\documentclass[conference]{IEEEtran}

\IEEEoverridecommandlockouts
\pdfoutput=1

\usepackage{cite}
\usepackage{amssymb,amsfonts}
\usepackage{amsmath}
\usepackage{indentfirst}
\usepackage{amsthm}
\usepackage{enumerate}

\usepackage{algorithm}
\usepackage{algorithmicx}
\usepackage{algpseudocode}

\usepackage{cleveref}

\usepackage{graphicx}
\usepackage{epstopdf}
\usepackage{subfigure}
\usepackage{subcaption}

\usepackage{booktabs}

\usepackage{textcomp}

\usepackage{xcolor}

\algnewcommand\BlueKeyword{\textcolor{blue}}

\newtheorem{lemma}{Lemma}
\newtheorem{theorem}{Theorem}

\floatname{algorithm}{Algorithm}

\crefname{figure}{figure}{figures}
\Crefname{figure}{Figure}{Figures}

\def\BibTeX{{\rm B\kern-.05em{\sc i\kern-.025em b}\kern-.08em
    T\kern-.1667em\lower.7ex\hbox{E}\kern-.125emX}}
\begin{document}

\title{TinyMA-IEI-PPO:  Exploration Incentive-Driven Multi-Agent DRL with Self-Adaptive Pruning for Vehicular Embodied AI Agent Twins Migration}
\author{Zhuoqi Zeng, Yuxiang Wei, Jiawen Kang* 
\thanks{
Y. Wei, Z. Zeng, J. Kang are with the Guangdong University of Technology, China (e-mail:
3122001501@mail2.gdut.edu.cn; 3123001489@mail2.gdut.edu.cn; kavinkang@gdut.edu.cn).

(\textit{*Corresponding author: Jiawen Kang}).
}
}

\maketitle

\begin{abstract}
 Embodied Artificial Intelligence (EAI) addresses autonomous driving challenges in Vehicular Embodied AI Networks (VEANETs) through multi-modal perception, adaptive decision-making, and hardware-software co-scheduling.  However, the computational demands of virtual services and the inherent mobility of autonomous vehicles (AVs) necessitate real-time migration of Vehicular Embodied Agent AI Twins (VEAATs) between resource-constrained Roadside Units (RSUs). This paper proposes a novel framework for efficient VEAAT migration in VEANETs, combining a multi-leader multi-follower (MLMF) Stackelberg game-theoretic incentive mechanism with a tiny multi-agent deep reinforcement learning (MADRL) algorithm. First, We propose an virtual immersive experience-driven utility model that captures AV-RSU dynamic interactions by integrating AVs' social influence, service complementarity and substitutability, and RSUs' resource allocation strategies to optimize VEAAT migration decisions. Second, to enhance training efficiency and enable efficient deployment on computation-constrained AVs while preserving exploration-exploitation performance, we propose  TinyMA-IEI-PPO, a self-adaptive dynamic structured pruning algorithm that dynamically adjusts neuron importance based on agents' exploration incentives. Numerical results demonstrate that our approach achieves convergence comparable to baseline models and closely approximates the Stackelberg equilibrium. 

\end{abstract}


\section{Introduction}
Rooted in Turing's embodied cognition theory, Embodied Artificial Intelligence (EAI) enables agents to interact with physical environments via sensorimotor coupling \cite{savva2019habitat}, emphasizing this coupling alongside situated intelligence to empower agents with perception, reasoning, and action capabilities in real-world contexts \cite{paolo2024call}. This is particularly evident in the integration of EAI with vehicular systems, which has led to the emergence of Vehicular Embodied AI Networks (VEANETs). In VEANETs, since Autonomous Vehicles (AVs) serve as the embodied agent, they are equipped with the ability to comprehensively perceive multimodal elements and make autonomous decisions \cite{sharma2024artificial}.

EAI bridges cyberspace and the physical world by integrating digital twins (DTs) to create Vehicle Embodied Agent Twins (VEATs) and Vehicle Embodied Agent AI Twins (VEAATs)  \cite{liu2024aligning} \cite{zhong2025generative}. VEATs leverage embodied simulators to virtual environments replicating real-world physics and synchronize physical-virtual spaces through real-time analytics\cite{zhang2025embodied}. These systems adopt embodied world models as their digital brain, integrating physics-aware reasoning with large-scale models such as multimodal large language models (MLLMs), large language models (LLMs), and vision-language models (VLM) \cite{xiang2023language}. VEAATs serve as AI assistants for in-vehicle application services in VEANETs \cite{zhong2025generative}. In VEANETs, the Fully Cognitive Intelligent Cockpit exemplifies VEAATs’ application in AVs \cite{li2023intelligent}, unifying hardware-software scheduling to enable real-time cabin occupancy perception, autonomous driving functions \cite{chen2024scenario}, and immersive infotainment via Head-Up Displays (HUDs).

However, the limited local computing resources of AVs pose challenges for executing and updating computation-intensive tasks in real-time. To address this, VEAAT tasks are offloaded to proximal ground base stations, such as Roadside Units (RSUs) equipped with edge servers, which provide adequate computing and bandwidth resources \cite{zhang2023learning}. RSUs supply computational resources for VEAAT task execution and allocate bandwidth for real-time VEAAT migration. However, the constrained RSU coverage and the constant mobility of AVs may cause AVs to progressively move away from their VEAATs \cite{chen2023multiagent}. To ensure continuous and dynamic interaction between the physical and virtual domains, VEAATs must undergo real-time migration from the current RSU to a new one. To achieve efficient and reliable VEAAT migration, we propose a MLMF game-theoretic incentive mechanism\cite{nie2020multi} \cite{li2019stackelberg}. This mechanism integrates the social influence of AVs, the strategic interconnections of RSUs, and a novel matching probability based on service immersion into the utility model.

Recent advancements in DRL have enabled algorithms to efficiently derive the Stackelberg Equilibrium (SE) in non-cooperative games while preserving the privacy of all players, making them suitable for complex multi-agent interaction scenarios \cite{Xu_Peng_Gupta_Kang_Xiong_Li_El-Latif_2022}. Meanwhile, Tiny Machine Learning (TinyML) and Few-Shot Learning (FSL) have become essential for resource-constrained environments. TinyML develops lightweight models via algorithmic approximation and pruning for embedded systems \cite{disabato2022tiny}, while FSL uses meta-learning to enable rapid task generalization from few labelled samples, reducing dependency on large annotated datasets \cite{song2023comprehensive}.

This paper primarily focuses on the innovative application of TinyML in multi-agent reinforcement learning, aiming to deploy lightweight MADRL models on computation-constrained AVs. High sample complexity persists as a major impediment to the application of DRL, especially in multi-agent systems \cite{loftin2021strategically}. To boost the efficiency of policy exploration and state data acquisition within complex state-space scenarios, we introduce individual exploration incentives as an intrinsic reward to encourage agents to explore behaviours that have a substantial impact on the global state. Furthermore, to preserve the exploration and exploitation performance of the model after lightweight, we propose a novel self-adaptive dynamic structured pruning method, termed Tiny \underline{M}ulti-\underline{A}gent \underline{I}ntrinsic \underline{E}xploration \underline{I}ncentive based \underline{P}roximal \underline{P}olicy \underline{O}ptimization (TinyMA-IEI-PPO). This algorithm adapts to changes in individual exploration incentives at different stages, dynamically adjusting pruning thresholds and formulating corresponding pruning strategies to gradually remove unimportant neurons with a binary mask.

To the best of our knowledge, this is the first work to integrate a self-adaptive dynamic structured pruning strategy, driven by individual exploration incentives, into the domain of DRL. The key contributions can be summarized as follows:

\begin{itemize}
    \item In VEANETs, constrained by limited RSU coverage and continuous vehicle mobility, we propose a VEAAT migration incentive mechanism based on a MLMF Stackelberg game. We define specific metrics for matching probability to capture the immersive experience from virtual service image quality and integrate the social influence of AVs and the complementarity and substitutability of SPs' services into the game's utility modelling.
    \item We innovatively enhance MAPPO’s training framework and objective function by introducing an intrinsic exploration mechanism that drives agents to prioritize actions with substantial impacts on global state transitions. 
    \item To balance model performance and neuron sparsity, we propose TinyMA-IEI-PPO, a tiny multi-agent deep reinforcement learning algorithm with self-adaptive dynamic structured pruning. It adapts to changes in individual exploration incentives during training, dynamically formulating pruning strategies. Results demonstrate the effective removal of redundant neurons while maintaining performance close to the Stackelberg Equilibrium.
\end{itemize}

The structure of the paper is organized as follows: Section \ref{Related} provides an overview of related works. Section \ref{System} introduces the system model, while Section \ref{stackelberg  game formulation and Equilibrium analysis for MLMF Stackelberg Game} delves into the Stackelberg Game formulation and the analysis of SE. Section \ref{tiny algorithm} elaborates on our proposed algorithm. The numerical results are presented in Section \ref{Results}, and Section \ref{Conclusion} concludes the paper.

\section{Related Works}\label{Related}
\subsection{Embodied AI in Vehicles}

In VEANETs, the integration of EAI into vehicular networks has emerged as a pioneering approach to address the complexities of autonomous driving and intelligent transportation systems. Similar to \cite{yang2024embodied}, EAI-enabled vehicles comprise two core components: a MLLM-based agent and an embodied entity. These components interact with both virtual and physical environments through modelling and sim2real operations, enabling real-time data gathering, processing, and feedback.

Zhou \textit{et al.} \cite{zhou2024embodied} proposed an embodied vision-language model with space-aware pre-training and time-aware token selection, enhancing agents' comprehension in long-range, dynamic environments. The authors in \cite{zhang2025embodied} combined LLM for semantic data processing with DRL for adaptive decision-making, optimizing real-time strategies in complex vehicular environments. A significant contribution is made in \cite{zhong2025generative}, which introduces the concept of EATs and VEAATs. These innovations collectively drive the realization of the Fully Cognitive Intelligent Cockpit in VEANETs \cite{li2023intelligent}, a paradigm that integrates hardware and software scheduling to elevate user experiences.VEAATs enhance the cockpit by supporting advanced autonomous driving features, alongside intelligent cabin monitoring and  VR/AR-based immersive infotainment systems.

\subsection{Resource Allocation Optimization in Twin Migration}

The establishment of virtual spaces and in-vehicle services demands significant resource consumption, driving resource allocation optimization in twin migration as a critical research focus. Recent studies have proposed innovative solutions to address key challenges. The authors in \cite{10608164} introduced an attribute-aware auction mechanism to optimize VT migration by considering monetary and non-monetary attributes. In \cite{10302973}, the authors leveraged a Stackelberg model with the "Age of Twin Migration" (AoTM) metric to promote efficient bandwidth allocation for rapid VT migration. Kang \textit{et al.} \cite{kang2024tiny} ensured real-time UAV Twins migration by incorporating a novel immersion metric, while \cite{10505943} addressed VT migration challenges through a multi-leader multi-follower game-theoretic incentive mechanism, integrating social awareness and queuing theory to optimize resource allocation. Notably, Zhong \textit{et al.} \cite{zhong2025generative} pioneered the consideration of twin migration within VEANETs, designing a Prospect Theory (PT)-based incentive mechanism to address VEAAT migration in uncertain environments by accounting for user preferences. However, none of these studies have considered the impact of the complementarity and substitutability of virtual services on users.

\subsection{DRL with Pruning Techniques and Related Advances}

Since traditional Heuristic and Meta-Heuristic Algorithms are limited in handling complex optimization problems, DRL emerges as a solution \cite{10734312}. DRL has proven to be a powerful tool for achieving equilibrium solutions in Stackelberg games. 
In MLMF Stackelberg games, agent strategies interact dynamically, with agents training distributed policies via historical local action observations as in Cooperative MARL. This setup faces inefficiencies like exponential joint action space growth, making effective exploration critical for maximizing cumulative rewards in complex environments \cite{kim2024strangeness}. MADRL exploration research has two main directions \cite{li2024individual}: global exploration (e.g., EMC \cite{zheng2021episodic} uses action value prediction errors for coordinated exploration rewards, though environmental dynamics may limit its effectiveness) and agent-level exploration (e.g., SMMAE \cite{zhang2023self} cultivates curiosity, and \cite{loftin2021strategically} designs efficient zero-sum game methods).

Nevertheless, training of DRL models also demands substantial computational resources and storage capabilities. To enhance the applicability of DRL models in resource-constrained scenarios, there is a significant demand for lightweight DRL solutions.
Prior works like policy distillation frameworks \cite{zhou2023online} and Policy Pruning and Shrinking (POPS) \cite{livne2020pops} have laid the foundational groundwork. Unstructured pruning typically yields irregular and non-compact network architectures, posing significant challenges for achieving effective training acceleration. Consequently, structured pruning is a more favorable alternative.Both the authors in \cite{kang2024tiny} and \cite{su2024compressing} adopt dynamic structured pruning with neuron importance group sparse regularization to penalize redundant neuron groups and gradually remove them. However, accurately evaluating neuron importance remains challenging due to the absence of definitive criteria. 

\begin{figure*}
    \centering
    \includegraphics[width=0.8\linewidth]{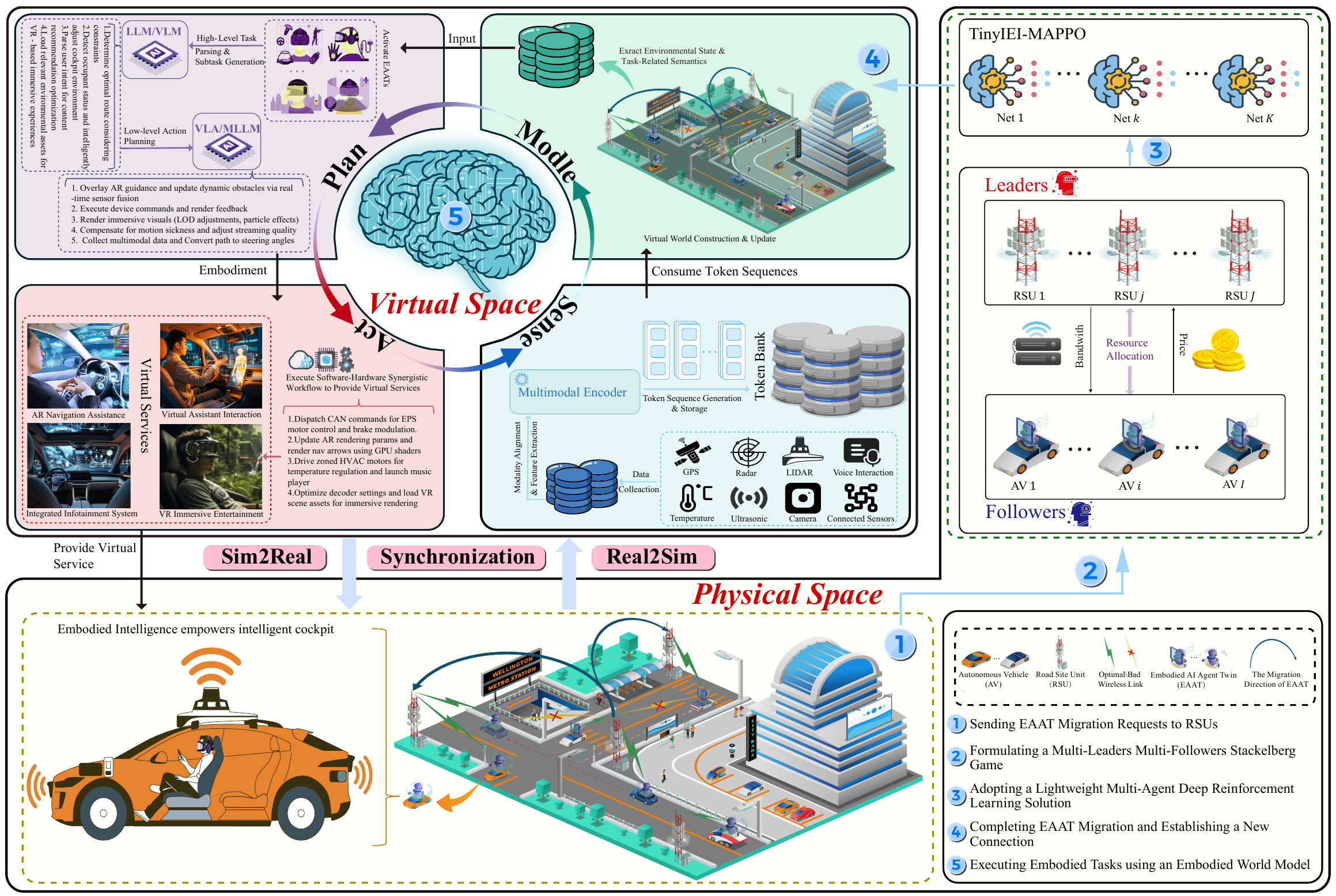}
    \caption{The system model for VEAAT migration. }
    \label{systemfig}
\end{figure*}

\section{System Model}\label{System}
\subsection{MLMF Stackelberg Game-based VEAAT Migration Framework}

Given the resource limitations of AVs, VEAATs are offloaded to RSUs for construction and updates. AVs upload real-time environmental data collected via sensors to RSUs to synchronize VEAATs in the virtual space, enabling RSUs to deliver corresponding services to in-vehicle users. To maintain uninterrupted service continuity amid AV mobility and limited RSU coverage, VEAATs must migrate dynamically between RSUs. AVs thus select optimal target RSUs by evaluating service types supported by their VEAATs, migration bandwidth requirements, and RSU pricing strategies, establishing a hierarchical resource management interaction framework between AVs and RSUs. Considering the interaction between RSUs and AVs, we propose a Stackelberg Game-based incentive mechanism framework between RSUs and AVs to optimize resource allocation during VEAAT migration. Our MLMF Stackelberg Game-based VEAAT Migration Framework and the steps for AVs to perform tasks in the context of intelligent cockpits are shown in Fig.\ref{systemfig}, and the detailed information is described as follows:

\textbf{Step 1. Sending VEAAT Migration Requests to Connected RSUs:} Before initiating VEAAT migration, the AV sends a migration request to the connected RSUs (i.e., the RSUs currently hosting the corresponding VEAATs and providing services). Subsequently, the connected RSUs broadcast the request to other RSUs and submit applications to these RSUs for purchasing bandwidth to support the VEAAT migration.

\textbf{Step 2. Formulating a MLMF Stackelberg Game:} After sending a VEAAT migration request, the AVs, as followers, must select which RSU to purchase bandwidth resources from and designate as the migration destination. In this selection process, RSUs act as leaders in the Stackelberg game, responsible for allocating bandwidth resources for VEAAT migration and independently determining pricing strategies for available bandwidth. AVs then decide the quantity of resource units to purchase for efficient VEAAT migration based on the bandwidth prices set by other RSUs. This interaction forms a MLMF Stackelberg game between RSUs and AVs for VEAAT migration.

\textbf{Step 3. Adopting a Lightweight MADRL Solution:} The TinyMA-IEI-PPO algorithm is deployed on the AVs to determine the optimal solution of the Stackelberg game. On the resource-constrained AVs, by deploying a lightweight MADRL model, strategies that meet the game equilibrium solution can be quickly and accurately formulated. The AVs select the optimal RSU for connection to carry out efficient and reliable VEAAT migration.

\textbf{Step 4. Completing VEAAT Migration Task and Establishing a New Connection:} The VEAAT is migrated from the original RSU to the target RSU. Upon arrival at the target RSU, it is added to the RSU's processing queue for re-instantiation. Once re-instantiation confirms the completion of the VEAAT migration task, AVs establish a new connection with the target RSU. The target RSU then continues to allocate resources for the VEAATs, ensuring seamless in-vehicle services for users in AVs.

\textbf{Step 5. Executing Embodied Tasks:} The process by which the embodied agents execute embodied tasks to provide services to users follows a \textbf{sense-model-plan-act} (SMPA) cycle \cite{CHRISLEY2003131}. \textbf{Sense:}  Embodied AI agents perceive environmental and self-states via multimodal sensors and user inputs, synchronizing this information into the virtual space constructed by connected RSUs. \textbf{Model:} Embodied agents integrate perceived data with prior knowledge to build and update dynamic environmental and self-state representations for corresponding EATs. \textbf{Plan:} Embodied task planning for agents integrates world models and MLLMs to first decompose abstract goals into executable subtasks through high-level reasoning, then generates software-hardware workflows via LLMs, VLMs, and Vision-Language-Action (VLA) models for real-time action sequencing. \textbf{Act:} The agents execute the planned actions according to the workflow for VEAAT service delivery. Meanwhile, they collect real-world service feedback and enter the SMPA cycle for adjustment and obtain rewards based on this feedback.

Our main work is to study steps 2 and 3, which use the MLMF Stackelberg game to summarize the interactions between AVs and RSUs, then use the TinyMA-IEI-PPO algorithm to determine the optimal solution to ensure efficient and reliable VEAAT migration while keeping the computational resource occupancy low.
\subsection{Total Delay Model of Migration Task}
In this paper, we consider VEAAT migration and resource trading involving multiple AVs and multiple RSUs in urban hotspots. Specifically, a set of $R$ RSUs and a set of $V$ AVs are represented by the set $\mathcal{R} = \{1,\ldots,j,\ldots,R\}, \mathcal{V}= \{1,\ldots,i,\ldots,V\}$, respectively. The VEAAT migration task of AV $i$ is denoted as $J_i = \{D_i, C_i, T^{\text{max}}_i, \alpha_i\}$, where $D_i$ is the total amount of migrated VEAAT data, including vehicle configuration, historical interaction data and real VEAAT state, $C_i$ is the number of CPU cycles required to re-instantiate VEAAT, $T^{\text{max}}_i$ is the maximum tolerated delay and $\alpha_i$ is used to describe their heterogeneity to measure the delay sensitivity of maintaining the virtual services supported by an VEAAT. Tasks can be divided into various types of tasks, such as AR navigation and the popular AR game \text{Pokémon} can be distinguished according to different $\{D_i, C_i, T^{\text{max}}_i, \alpha_i\}$. 

In hot spot areas, dense building layouts induce non-line-of-sight (NLoS) propagation with Rayleigh-distributed fading, which is addressed by deploying reconfigurable intelligent surfaces (RIS) to optimize multipath signals through dynamic phase-amplitude adjustments \cite{10352433}. Meanwhile, multi-user MIMO enhances system capacity by leveraging spatially independent channels for concurrent multi-AV data streaming \cite{wang2025pre}. To sum up, for a given bandwidth $b_{ij}$ purchased from $RSU_j$ by $AV_i$ and considering the Rayleigh fading channel, the transmission data rate can be calculated as $r_{ij}=b_{ij}\log_2\left(1+\frac{\rho h_{ij}}{\sigma^2}\right)$, where $\rho$ represents the transmitter power of AV, and $\sigma^2$ signifies the power of the additive white Gaussian noise (AWGN) \cite{10302973}, and $h_{ij} = A\left(l/{4\pi fd_{ij}}\right)^2$ denotes the magnitude of the channel gain that follows a Rayleigh distribution, where $A$ denotes the channel gain coefficient, $l$ is the speed of light, $f$ denotes the carrier frequency and $d_{ij}$ is the Euclidean distance between AV $i$ and RSU $j$ \cite{chen2023multiagent}.

The total migration delay contains three parts \cite{10505943}: the transmission delay, the queuing delay and the re-instantiation delay. First, the transmission delay from $AV_i$ to $RSU_j$ can be expressed as $t_{ij}^\mathrm{tran}=\frac{D_{i}}{r_{ij}}$. 
Second, the queuing delay $t_{ij}^\mathrm{que}$ arises from processing congestion RSU $j$, modeled using an M/M/1 queue \cite{li2019stackelberg} with arrival rate $\lambda_j$ and server processing rate $\mu_j$, expressed as \(t_{ij}^\mathrm{que}=\frac{\lambda_j}{\mu_j(\mu_j - \lambda_j)}\). Finally, the re-instantiation delay $t_{ij}^\mathrm{com}$ reflects the computational time for rebuilding the VEAAT at RSU $j$, given by $t_{ij}^\mathrm{com}=\frac{C_{i}}{f_{j}}$. Consequently, the total migration delay of task $J_i$ is denoted as $T_{ij}=t_{ij}^\mathrm{tran}+t_{ij}^\mathrm{que}+t_{ij}^\mathrm{com}.$

\subsection{Utility Modelling of AVs and RSUs in Stackelberg Game}
The interaction between AVs and RSUs can be modeled as a two-stage Stackelberg game framework. RSUs are resource providers targeting consumers, and they announce the prices at which they sell bandwidth resources to AVs for VEAAT migration tasks \cite{li2019stackelberg}, while AVs pay the RSUs and obtain bandwidth resources from the RSUs. The selling prices of all RSUs are defined as the vector $\boldsymbol{P}=\left\{p_j\right\}_{j\in\mathcal{R}}$, and the bandwidth requirements of all AVs are defined as the vector $\boldsymbol{B}=\left\{b_i\right\}_{i\in\mathcal{V}},$where $b_{ij}=\left\{b_{i}\right\}_{j\in\mathcal{R}}$ represents the vector of bandwidth purchased from all RSUs.

Inspired by prior research \cite{Huang_Zhong_Nie_Hu_Xiong_Kang_Quek_2022}, we assume that each AV adopts a probabilistic decision-making model to choose an RSU for purchasing bandwidth resources. However, the matching probabilities considered in these works solely rely on the relationship with resource pricing. This simplistic approach fails to comprehensively evaluate the quality of resource services offered by RSUs and fully capture the true matching preferences of users within AVs. Notably, in \cite{kang2024tiny}, the authors introduced a novel metric, "Meta-Immersion", to estimate the quality of experience (QoE) that Unmanned Aerial Vehicles Metaverse Users experience in virtual services. 
Without loss of generality, we adapt this metric to our scenario: as EAI-empowered AVs handle driving tasks autonomously, passengers prioritize in-vehicle entertainment, where immersion hinges on AR/VR image quality enabled by VEAATs. Our paper takes the QoE into account and redefines the connotation of the matching probability between the  RP $j$ and the SP $i$, which is expressed as 

\begin{equation}
    \theta_{ij}=\frac{\frac{q_j}{P_j}}{\Sigma_{l\in\mathcal{R}}\frac{q_l}{P_l}},
\end{equation}
    where $q_j$ denotes the QoE that an AV perceives from the services provided by  RSU $j$. In \cite{ding2021locally}, the authors proposed the locally adaptive DISTS metric, namely A-DISTS, which is a new full-reference image quality assessment (IQA) metric. When seeking an immersive user experience in virtual services, real-time rendering becomes a key technology, with graphics rendering as the main function \cite{yu2024attention}. The fidelity of image rendering plays a crucial role in shaping the immersive experience. We use the A-DISTS metric instead of the SSIM metric to solve the problem of ignoring the local structure and texture features of images in full-reference IQA and take $A\mathrm{-}DISTS_j$ as the user's quality rating for the graphic rendering in the service provided by RSU $j$. The calculation formula of $A\mathrm{-}DISTS_j$ is as follows:
\begin{equation}
A\mathrm{-}DISTS_j(X,Y)=1-\frac{1}{Z}\sum_{c=0}^C\sum_{z=1}^{Z_c}S(\tilde{X}_z^{(c)},\tilde{Y}_z^{(c)}),\end{equation}
where $C$ represents the number of convolutional stages, $Z_c$  is the number of feature maps at the $c\mathrm{-}$th convolutional stage, $Z$$=$$\sum_{c=0}^CZ_c$ and $S(\tilde{X}_z^{(c)},\tilde{Y}_z^{(c)})$ is used to compute the similarity between the predicted image $\tilde{X}_z^{(c)}$ and the reference image $\tilde{Y}_z^{(c)}$ on the $z\mathrm{-}$th feature map of the $c\mathrm{-}$th stage. The specific calculation formula is
\begin{equation}
S(\tilde{X}_z^{(c)},\tilde{Y}_z^{(c)})=\frac{1}{Q_c}\sum_{k=1}^{Q_c}\left(\tilde{p}_k^{(c)}l(\tilde{x}_{z,k}^{(c)},\tilde{y}_{z,k}^{(c)})+\tilde{q}_k^{(c)}s(\tilde{x}_{z,k}^{(c)},\tilde{y}_{z,k}^{(c)})\right).
\end{equation}
In the formula provided, $Q_c$ denotes the number of local regions calculated on the feature map at the $c\mathrm{-}$th stage.The terms $\tilde{p}_k^{(c)}$ and $\tilde{q}_k^{(c)}$ represent the texture probability of the $k\mathrm{-}th$ patch observed at the $c\mathrm{-}$th cale and its complement, respectively. Similar to SSIM, the functions ${{l}}(\cdot)$ and ${{s}}(\cdot)$ are specifically defined to quantify the structural and textural similarities. Therefore, the QoE level $q_j$ based on Weber-Fechner's Law (WFL) \cite{Du_Liu_Niyato_Kang_Xiong_Zhang_Kim_2022} that RSU $j$ can provide to users within AVs can be expressed as  
\begin{equation}
 q_j=k_j\ln{(\frac{A\mathrm{-}DISTS_j}{A\mathrm{-}DISTS_j^{th}})}, 
\end{equation}
where $A\mathrm{-}DISTS_j^{th}$ represents the users' minimum image rendering quality requirement, and $k_j$ is the service reputation weight parameter, which increases with the number of times the RSU has provided high-quality virtual service rendering and completed VEAAT migration tasks.
\subsubsection{Utility Function of AVs}
We focus on the positive social network effects among AVs as well as the complementarity and substitutability of various service applications provided by RSUs \cite{li2019stackelberg} \cite{nie2020multi}. Consequently, while ensuring the non-cooperative game relationship among absolutely rational individuals in the game model, we have appropriately incorporated some external rewards to more accurately reflect the actual perceptual effects of the services as experienced by users. The utility of AV $i$ is then defined as follows:
\begin{equation}
    U_{i}^{F}=V_{i}(b_{i})+\Phi_{i}(b_{i},B_{-i}) 
+\beta_{i}(b_{i})+C_{i}(b_{i}).
\end{equation}

The first term $V_{i}(b_{i})$ indicates the fact that each AV $i$ can obtain internal benefits from the participation in all RSUs. To more precisely model the connection between human perception and relative stimulus changes, we adopt the natural logarithm function $\ln(\cdot)$ to model the internal benefits based on WFL.  This approach not only ensures the convexity of the utility function but also accurately captures the nonlinear traits of perception. As a result, the first term $V_{i}(b_{i})$ is denoted as follows: 
\begin{equation}
V_{i}(b_{i})=\sum_{j\in \mathcal{R}}\theta_{ij}[\delta_{i}\ln(b_{ij}+e-d_{i}T_{i}^{\max})],
\end{equation}
where $\delta_{i}$ is the maximum internal satisfaction factor, $d_{i}$ is the maximum delay tolerance factor, and $e$ is the natural constant.

We incorporate the second term, $\Phi_{i}(b_{i},B_{i})$, to represent the external benefits gained from the positive social network effects among AVs. Specifically, when AVs increase and purchase more bandwidth resources for their services, it has a positive impact on other AVs. For example, in the scenario of VEAATs involved in the multi-player AR game NFT All-Stars, when an AV increases its bandwidth procurement for its associated VEAAT, the VEAATs on other RSUs can share network bandwidth through resource-sharing nodes. This optimizes the network in the decentralized infrastructure, enabling more efficient rendering of the virtual world and players' blockchain-based NFT avatars. The sharing reduces resource consumption at the destination, ultimately enhancing the service experience for all participating AVs \cite{10505943}. To model network effects, we introduce the adjacency matrix $\boldsymbol{G_1}=[\zeta_{ik}]$, where $ i,k\in \mathcal{V}$. The element $\zeta_{ik}$ in the  $i\mathrm{-}$th row and  $k\mathrm{-}$th column of the matrix $\boldsymbol{G}$ represents the mutual social influence between AV $i$ and AV $k$ (i.e., $\zeta_{ik}=\zeta_{ki}$). The second term $\Phi_{i}(b_{i},B_{i})$ is denoted as follows:
\begin{equation}
    \Phi_{i}(b_{i},B_{i})=\sum_{j\in \mathcal{R}}\theta_{ij}\sum_{k\in \mathcal{V}\setminus i}\zeta_{ik}b_{kj}b_{ij}.
\end{equation}

The third term is \begin{equation}
    \beta_{i}(b_{i},B_{i})= \sum_{j\in \mathcal{R}}\theta_{ij}\sum_{s\in \mathcal{R}\setminus j}\eta_{js}b_{ij}b_{is},
\end{equation} which captures the complementarity and substitutability of service applications that are offloaded to RSUs along with VEAATs. The parameter $\eta_{is}$ represents the interconnection between RSU $j$ and $s$. Here, we still introduce the adjacency matrix $\boldsymbol{G_2}=[\eta_{is}]$ and assume $\eta_{is}=\eta_{si}$, $\text{for all }j,s\in \mathcal{R}.$ When VEAATs are offloaded to RSU $j$ for Simultaneous Localization and Mapping (SLAM) services, while another RSU $s$ supports Pokémon Go through its VEAATs, a complementary relationship emerges. Specifically, RSU $s$ can leverage the SLAM API provided by RSU $j$, which utilizes real-time vehicle data such as angular velocity and acceleration to construct precise environmental models for autonomous vehicles. This cross-RSU collaboration enhances the integration of Pokémon characters and improves positioning/tracking accuracy in the augmented reality game. The positive value of the complementarity indicator $\eta_{is}>0$ reflects this synergistic effect between the two services.

In contrast, due to the presence of competitive application services, take AR navigation as an example. We assume that market demand is symmetrical, meaning that market segmentation among different RSUs in AR navigation is likely to be evenly distributed. In the absence of a dominant application service with overwhelming market power, there are multiple viable options, such as AutoNavi, Tencent Maps, and Google Maps. If the VEAAT offloads to RSU $j$ opts to support AutoNavi in the AR-navigation service, while RSU $s$ supports Google Maps, due to service substitutability, RSU $s$ will retain resource occupancy and refuse to share or cooperate with RSU $j$. This behavior will have a negative impact on RSU $j$. Consequently, the service experience of the AV associated with the VEAAT will decline. In this case, $\eta_{is}<0$, indicating that the services of  RSU $s$ and RSU $j$ are substitutable.

Following standard practice, the AV pays the RSU post-VEAAT migration as per the Service Level Agreement (SLA). Consequently, for the fourth term, we define the cost function $C_i(b_i)$ as $\sum_{j\in \mathcal{R}}\theta_{ij}p_{ij}b_{ij}$. To sum up, the utility of AV $i$ is formulated as follows:

\begin{equation}
    \begin{split}
        U_i^F(b_i, \boldsymbol{B}_{-i}, \boldsymbol{P}) 
        &=  \sum_{j \in \mathcal{R}} \frac{\frac{q_j}{P_j}}{\sum_{l \in \mathcal{R}} \frac{q_l}{P_l}} \left[ \delta_i \ln(b_{ij} + e - \alpha_i T_i^{\text{th}}) \right. \\
        &\left. + \sum_{k \in \mathcal{V} \setminus i} \zeta_{ik} b_{ij} b_{kj} + \sum_{s \in \mathcal{R} \setminus j} \eta_{js} b_{ij} b_{is} - p_{j} b_{ij} \right].
    \end{split}
    \label{Ulity of F}
\end{equation}

\subsubsection{Utility Function of RSUs}
RSUs are required to allocate adequate bandwidth resources to AVs for VEAAT migration, which entails associated costs such as those for transmission and re-instantiation. The cost for RSU $j$ to provide bandwidth resources to AV $i$ for the corresponding VEAAT migration is denoted as $c_{ij}$.
Given the pairing probability $\theta _{ij}$, the Utility Function (i.e., the profit) of each RSU is calculated as the difference between the total bandwidth fees paid by the paired AVs and the cost of processing the VEAAT migration task. Thus, we can define the utility function for RSU $j$ as follows:

\begin{equation}
U_{j}^{L}(p_{j}, \boldsymbol{P}_{-j}, \boldsymbol{B}) = \sum_{i \in \mathcal{I}} \frac{\frac{q_{j}}{p_{j}}}{\sum_{l \in \mathcal{R}} \frac{q_{l}}{p_{l}}} \left[ b_{ij} (p_{j} - c_{ij}) \right].
\label{Ulity of L}
\end{equation}

\section{stackelberg game formulation and Equilibrium analysis for MLMF Stackelberg Game}\label{stackelberg  game formulation and Equilibrium analysis for MLMF Stackelberg Game}

\subsection{Stackelberg Game Formulation}
 RSUs and AVs are considered to be absolutely rational and self-interested, needing to independently determine the optimal bandwidth purchase strategies and selling prices to maximize their respective utilities. Therefore, we formulate the interaction between RSUs and AVs as a MLMF Stackelberg game with two stages. In the Stackelberg game, players consist of leaders and followers. In stage \text{I}, the leaders (RSUs) first decide on their selling prices for bandwidth resources. Subsequently, in stage \text{II}, the followers (AVs) then determine their bandwidth requirements accordingly based on their VEAAT migration tasks and the unit selling prices of bandwidth from RSUs. In the follower sub-game, given the distribution of bandwidth prices from all RSUs $P$ and the bandwidth demand distribution of AVs (i.e., ${B_{-i}}$), the AV $i$'s objective is to maximize its own utility by solving the following optimization problem: 
\begin{equation}
    \begin{split}
    \textbf{\textit{P1:}}\:&\max\limits\: U_i^F(b_i, \boldsymbol{B_{-i}}, \boldsymbol{P}),  \\
    &\:\:s.t.\:\: {b_{ij} \geq 0},\\
    &\quad\:\:\:\:\: \sum_{j\in\mathcal{R}}{\theta_{ij}}{T_{ij} \leq T^{max}_i}.
    \end{split}
    \label{U_F}
\end{equation}

In the leader sub-game, the RSU's objective is to optimize its utility based on the pricing strategies adopted by all other RSUs (i.e., $\boldsymbol{P}_{-j}$) and the bandwidth purchasing strategies employed by all AVs. The specific optimization problem is formulated as follows:
\begin{equation}
    \begin{split}
    \textit{\textbf{P2:}}\:&\max\limits\:U^L_j(p_j,{\boldsymbol{P^*_{-j}}},\boldsymbol{B}),  \\
    &\:\:s.t.\:\: {p_{ij}\in[c_j, p^{max}]},
    \end{split}
    \label{U_L}
\end{equation}

where $p^{max}$ represents the upper limit of the selling price $p_j$. This constraint reflects that both RSUs and AVs are absolutely rational and self-interested entities. When RSU $j$ sets its selling price above the AVs' expectations, no AVs would be willing to pay for the bandwidth. At the same time, RSUs will not price their bandwidth resources too low in an attempt to increase the likelihood of AVs purchasing their services, as doing so would result in a loss.

In a Stackelberg equilibrium, all players, including RSUs and AVs, aim to maximize their individual payoffs during the decision-making process. The Stackelberg equilibrium is defined as a stable point where the leaders' payoffs are optimized, given that the followers have adopted their optimal strategies \cite{10302973}. The Stackelberg equilibrium can be defined as follows:

\textbf{Definition 1}: (Stackelberg Equilibrium, SE): \textit{The optimal bandwidth demand strategies and the optimal bandwidth selling prices are denoted as$\boldsymbol{B^\ast}=\left\{b_i^\ast\right\}_{i\in\mathcal{V}}$ and $\boldsymbol{P^\ast}=\left\{p_j^\ast\right\}_{j\in\mathcal{R}}$, respectively. Let $\boldsymbol{B^\ast_{-i}}$ represent the optimal bandwidth demand strategies of all other AVs except for $i$, and $\boldsymbol{P^\ast_{-j}}$ represent the optimal bandwidth selling price strategies of all other RSUs except for $j$. Then, the stable point $(\boldsymbol{B^\ast},\boldsymbol{P^\ast})$ is denoted as a SE, where RSUs and AVs cannot increase their profit by changing their strategies, that is, when the following inequalities are strictly satisfied\cite{Huang_Zhong_Nie_Hu_Xiong_Kang_Quek_2022}: 
\begin{equation}
\begin{cases}
U_i^F\big(b_i^*,\boldsymbol{B^\ast_{-i}},\boldsymbol{P^\ast})\geq U_i^F\big(b_i,\boldsymbol{B^\ast_{-i}},\boldsymbol{P^\ast}),\forall i\in\mathcal{V}, \\
U_j^L\big(p_j^*,\boldsymbol{P^\ast_{-j}},\boldsymbol{B^\ast})\geq U_j^L(p_j,\boldsymbol{P^\ast_{-j}},\boldsymbol{B^\ast}),\forall j\in\mathcal{R}.
\end{cases}
\end{equation}}

\subsection{Stackelberg Equilibrium Analysis}

In this section, we employ backward induction to study the Stackelberg equilibrium. We first analyze the non-cooperative game at the follower level in Stage II by finding the Nash equilibrium solutions and proving their existence and uniqueness given the strategies of the RSUs. Subsequently, we substitute the Nash equilibrium solutions of the follower-level game into the leader-level non-cooperative game in Stage I and further demonstrate the existence and uniqueness of the Stackelberg equilibrium.
\subsubsection{Analysis of the Follower-level Game}

 In stage I, every AV \(i\) modifies its bandwidth requirement. The objective is to maximize its utility, and this adjustment is made according to the price profiles \(\boldsymbol{P}\) of all RSUs. For the convenience of subsequent calculations and proofs, we set \(y_j=\frac{1}{p_j}\), set \(\delta_{i}(1+\sum_{k\in U\setminus i}\zeta_{ik}b_{kj}+\sum_{s\in R\setminus j}\eta_{js}b_{is})\) as $A$, and $e-\alpha_{ij}T^{th}_i$ as $E$. \\

\begin{lemma}
\textit{The existence of Nash equilibrium in a non-cooperative game can be guaranteed when the following three conditions are met \cite{415dc80a-289c-39e1-8e6b-601fc5ef267e}: 
\begin{itemize}
\item The player set is characterized by finiteness.
\item Both strategy sets are delineated by closure and boundedness, demonstrating convexity.
\item The utility functions exhibit continuity and quasi-concavity within the confines of the strategy space.
\end{itemize}} 
\label{lemma 1}
\end{lemma}

\begin{theorem}
\textit{There exists a Nash equilibrium in the non-cooperative game among RSUs.}     
\end{theorem}

\textit{Proof.} The first-order and second-order derivatives of \(U_{i}^{F}\) with respect to \(b_{ij}\)  are derived as follows:

\begin{equation}
    \begin{split}
        \frac{\partial U_{i}^{F}}{\partial b_{ij}}&=\sum_{j\in \mathcal{R}}\frac{\frac{q_{j}}{p_{j}}}{\sum_{l\in \mathcal{R}}\frac{q_{l}}{p_{l}}}\left[\frac{A}{b_{ij}+E}-p_{j}\right].
    \end{split}
    \label{eq:derivative} 
\end{equation}

\begin{equation}
    \begin{split}
        \frac{\partial^{2} U_{i}^{F}}{\partial b_{ij}^{2}}&=\sum_{j\in \mathcal{R}}\frac{\frac{q_{j}}{p_{j}}}{\sum_{l\in \mathcal{R}}\frac{q_{l}}{p_{l}}}\left[\frac{-A}{(b_{ij}+E)^{2}}\right] <0.
    \end{split}
    \label{eq:derivative2} 
\end{equation}
The negative second-order derivative presented in Eq.(\ref{eq:derivative2}) implies the quasi-concavity of the utility function $U^{F}_{i}$ with respect to ${b_i}$. Then, by applying the first-order optimality condition \( \frac{\partial U_i^F}{\partial b_{ij}} = 0 \), we can derive the optimal strategy of AV \( i \) towards RSU \( j \) as follows:
\begin{equation}
b_{ij}^*=\frac{A}{p_{ij}}-E.
\label{b_bestresponse}
\end{equation}
In addition, the strategy set of AVs satisfies the basic criteria of being closed, bounded, and convex. Moreover, considering the finite nature of the AV set and the continuity of its utility function, it can be considered that the Nash equilibrium among AVs exists according to Lemma \ref{lemma 1}.
If the best response function of the AV conforms to the standard form, then a unique Nash equilibrium exists in the follower sub-game \cite{Xu_Qiu_Zhang_Liu_Liu_Chen_2021}.

\begin{lemma}
\label{Lemma 2} \textit{A function $\mathcal{X}(\boldsymbol{B})$ is a standard function if and only if it satisfies the following three conditions:
\begin{itemize}
    \item Positiveness: $\mathcal{X}(\boldsymbol{B})>0$
    \item Monotonicity: $\forall\boldsymbol{B'}>\boldsymbol{B},\mathcal{X}(\boldsymbol{B'})>\mathcal{X}(\boldsymbol{B})$
    \item Scalability: $\forall x>1,x\mathcal{X}(\boldsymbol{B})>\mathcal{X}(x\boldsymbol{B})$
\end{itemize}}
\end{lemma}

\begin{theorem}
\textit{If $p_{j}<\frac{\delta_{i}}{\alpha_{i}T_{i}^{th}-e}$ is satisfied, the sub-game perfect equilibrium in the AVs' sub-game is unique.}   
\end{theorem}

\textit{Proof.} Let $\varphi_{ij}=\frac{A}{p_{ij}}-B$. Then, we further obtain the best-response function as follows:
\begin{equation}
b_{ij}^*=\mathcal{X}_{ij}(\boldsymbol{B})=
\begin{cases}
0, & \psi_{ij}<0, \\
\psi_{ij}, & \psi_{ij}\geq0.
\end{cases}
\label{best b}
\end{equation}

According to Lemma \ref{Lemma 2}, if the best-response function given in Eq.(\ref{best b}) satisfies Positiveness, Monotonicity, and Scalability, we can prove the uniqueness of the sub-game of AVs. First, these three properties are satisfied at the lower bound, i.e.,  $\mathcal{X}_{ij}(\boldsymbol{B})=0, (\psi_{ij}<0)$. Then we analyze $\mathcal{X}_{ij}(\boldsymbol{B})=\psi_{ij}$. We consider that the negative impacts brought by mutual substitutability are smaller than the positive impacts brought by complementarity and social network effects, i.e., $\sum_{k\in\mathcal{V}\setminus i}\zeta_{ik}b_{ij}b_{kj}+\sum_{s\in\mathcal{R}\setminus j}\eta_{js}b_{ij}b_{is}>0$. For positiveness, we can easily get $\psi_{ij}>0$; secondly, for monotonicity, let $\boldsymbol{B'}>\boldsymbol{B}$, then there exists $\sum_{k\in\mathcal{V}\setminus i}\zeta_{ik}b^{'}_{kj}+\sum_{s\in\mathcal{R}\setminus j}\eta_{js}b^{'}_{is}>\sum_{k\in\mathcal{V}\setminus i}\zeta_{ik}b_{kj}+\sum_{s\in\mathcal{R}\setminus j}\eta_{js}b_{is}$. Thus, it is easy to prove the monotonicity condition. For scalability, it can be proved from Eq.(\ref{eq:scblability 1}) that it satisfies scalability. In conclusion, the best-response function of AVs satisfies the three characteristic properties of the standard function. Therefore, we have proved the existence and uniqueness of the follower-level Nash equilibrium. The best-response function of AVs indicates that the higher the bandwidth price set by RSU $j$, the less amount of bandwidth is purchased by AV $i$.
\begin{equation}
\begin{split}
    &x\mathcal{X}_{ij}(\boldsymbol{B})-\mathcal{X}_{ij}(x\boldsymbol{B})=\\
    &\frac{xA}{p_{j}}-
        \frac{\delta_{i}\left[1+\sum_{k\in \mathcal{V}\setminus i}x\zeta_{ik}b_{kj}+\sum_{s\in \mathcal{R}\setminus j}x\eta_{js}b_{is}\right]}{p_{j}}+(xE\\&-E)=(x-1)(\frac{\delta_i}{p_{j}}-E)>0
\end{split}
\label{eq:scblability 1}
\end{equation}
\subsubsection{Analysis of the Leader-level Game}
\begin{theorem}
The unique Stackelberg Equilibrium exists in the MLMF Stackelberg game between RSUs and AVs, where both the bandwidth-demand strategies of AVs and the bandwidth-price strategies of RSUs are optimized.
\end{theorem}
\textit{Proof.} After each follower selects the optimal bandwidth-demand strategy, RSUs can maximize their utility by adjusting the optimal \(p_j\).Then the optimal strategies of AV \(i\), as given in Eq.(\ref{b_bestresponse}), are substituted into the utility function of RSU \(j\) as follows:
\begin{equation}
\begin{split}
&U_{j}^{L}(p_{j},\boldsymbol{P}_{-j},\boldsymbol{B}) = \\
&\sum_{i\in \mathcal{V}}\frac{1}{\sum_{i\in \mathcal{R}}q_{i}y_{i}}(-Aq_{j}c_{j}y_{j}^{2}+Aq_{j}y_{j}+Eq_{j}c_{j}y_{j}-Eq_{j}).
 \end{split}
\end{equation}

By computing the first and second-order derivative of \(V_j\) with respect to \(p_{j}\), the following expressions are derived, i.e.,
\begin{equation}
    \begin{split}
        \frac{\partial U_j^L}{\partial y_j} &= \sum_{i\in \mathcal{V}}\frac{1}{(\sum_{l\in \mathcal{R}}q_ly_l)^2} \left[(-2Aq_jc_jy_j+Aq_j+Eq_jc_j)\right.\\
        &\quad\left.\cdot\sum_{l\in \mathcal{R}\setminus j}q_ly_l-Aq_j^2c_jy_i^2+Eq_j^2\right].
    \end{split}
\end{equation}
\begin{equation}
    \begin{split}
    \frac{\partial^2 U_j^L}{\partial y_j^2} &=\sum_{i\in \mathcal{V}}\frac{-2q_j}{(\sum_{l\in \mathcal{R}}q_ly_l)^3}\left[c_jA\left(\sum_{l\in \mathcal{R}\setminus j}q_ly_l\right)^2\right.\\
        &\quad\left.+(Aq_j+Eq_jc_j)\cdot\sum_{l\in \mathcal{R}\setminus j}q_ly_l+Eq_j^2\right]<0.
    \end{split}
\end{equation}
The calculation results show that the second-order derivative is strictly negative. By setting the first-order derivative of \(U^L_j\) as 0 and considering the upper and lower limits of the set price, we can obtain the optimal strategy of RSU \(j\) for AV \(i\) expressed as
\begin{equation}p_j^*=\mathcal{H}_j(\boldsymbol{Y})=
\begin{cases}
0, & \omega_j<c_j ,\\
\omega_j, & 0\leq\omega_j\leq p^{max} ,\\
p^{max}, & \omega_j>p^{max},
\end{cases}\end{equation}
where\begin{equation}
\begin{split}
     &\omega_j=\\
 &\sum_{i\in \mathcal{V}}\frac{\sum_{l\in \mathcal{R}\setminus j}q_{l}y_{l}-\sqrt{(\sum_{l\in \mathcal{R}\setminus j}q_{l}y_{l}+\frac{q_{j}}{c_{j}})(\sum_{l\in \mathcal{R}\setminus j}q_{l}y_{l}+\frac{Eq_{j}}{A})}}{-q_{j}}.
\end{split}
 \end{equation}Similar to the analysis of the Follower-level game, in order to establish the uniqueness of the Nash equilibrium at the Leader-level, we also need to check whether the best-response function of RSU satisfies the three properties mentioned in Lemma \ref{Lemma 2}.

 It is obvious that the conditions of the standard function are met in other cases. Therefore, we only need to conduct an analysis when \(0 \leq \omega_j \leq p_{\max}\).
For positivity, it is easy to observe that the value inside the square root is greater than the value outside the square root. Thus, it satisfies this property. For monotonicity, according to the chain rule of differentiation, we can deduce that: $\frac{\partial H_{j}(y)}{\partial y}=\frac{\partial H_{j}(y)}{\partial(\sum_{l \in \mathcal{R}\setminus j}q_ly_{l})}\cdot\frac{\partial(\sum_{l \in \mathcal{R}\setminus j}q_ly_{l})}{\partial y}$.  Let $\sum_{l \in \mathcal{R}\setminus j}q_ly_{l}$ be \(G\), and then we can derive
\begin{equation}
\begin{split}
    &\frac{\partial H_{j}(\boldsymbol{Y})}{\partial G}=\frac{\partial(\sum_{i\in \mathcal{V}} \frac{G-\sqrt{(G+\frac{q_{i}}{G})(G+\frac{Eq_{j}}{A}}}{-g_{j}})}{\partial G}=-\frac{1}{q_{j}}+\\
    &\frac{G+\frac{q_{j}}{2}(\frac{1}{c_{j}}+\sum_{i\in \mathcal{V}}\frac{E}{A})}{q_{j}\sqrt{[G+\frac{q_{j}}{2}(\frac{1}{c_{j}}+\sum_{i\in \mathcal{V}}\frac{E}{A})]^{2}-\frac{q_{j}^{2}}{4}(\frac{1}{c_{j}}-\sum_{i\in \mathcal{V}}\frac{E}{A})^{2}}}\geq0.
\end{split}\end{equation}

When \(\boldsymbol{Y'}\geq\boldsymbol{Y}\), then \(\sum_{l \in \mathcal{R} \setminus j} q_ly_l' \geq \sum_{l \in \mathcal{R}\setminus j}  q_ly_l\).
Also, since \(\frac{\partial \sum_{l \in \mathcal{R}(j)} y_l \cdot q_l}{\partial y}>0\), to prove \(\frac{\partial H_{j}(\boldsymbol{Y})}{\partial y} \geq 0\), it is only necessary to prove \(\frac{\partial H_{j}(\boldsymbol{Y})}{\partial G} \geq 0\).

For scalability, it can be proved by the following formula:
\begin{equation}\begin{aligned}
 & \lambda\mathcal{H}_{j}(\boldsymbol{Y})-\mathcal{H}_{j}(\lambda\boldsymbol{Y}) = \\
 &\left(\sum_{i\in \mathcal{V}}\frac{\lambda G-\sqrt{(\lambda G+\frac{\lambda q_{j}}{c_{j}})(\lambda G+\frac{\lambda Eq_{j}}{A})}}{-q_{j}}\right) \\
 & -\left(\sum_{j\in \mathcal{V}}\frac{\lambda G-\sqrt{(\lambda G+\frac{q_{j}}{c_j})(\lambda G+\frac{Eq_{j}}{A})}}{-q_{j}}\right)>0.
\end{aligned}\end{equation}
Therefore, we can prove that the best-response function of RSU adheres to a standard function, ensuring that there is a unique Nash equilibrium in the leader-level subgame. In conclusion, we finally affirm that a Stackelberg Equilibrium exists and is unique in the formulated MLMF Stackelberg game between RSUs and AVs.

\begin{figure*}
    \centering
    \includegraphics[width=0.8\linewidth]{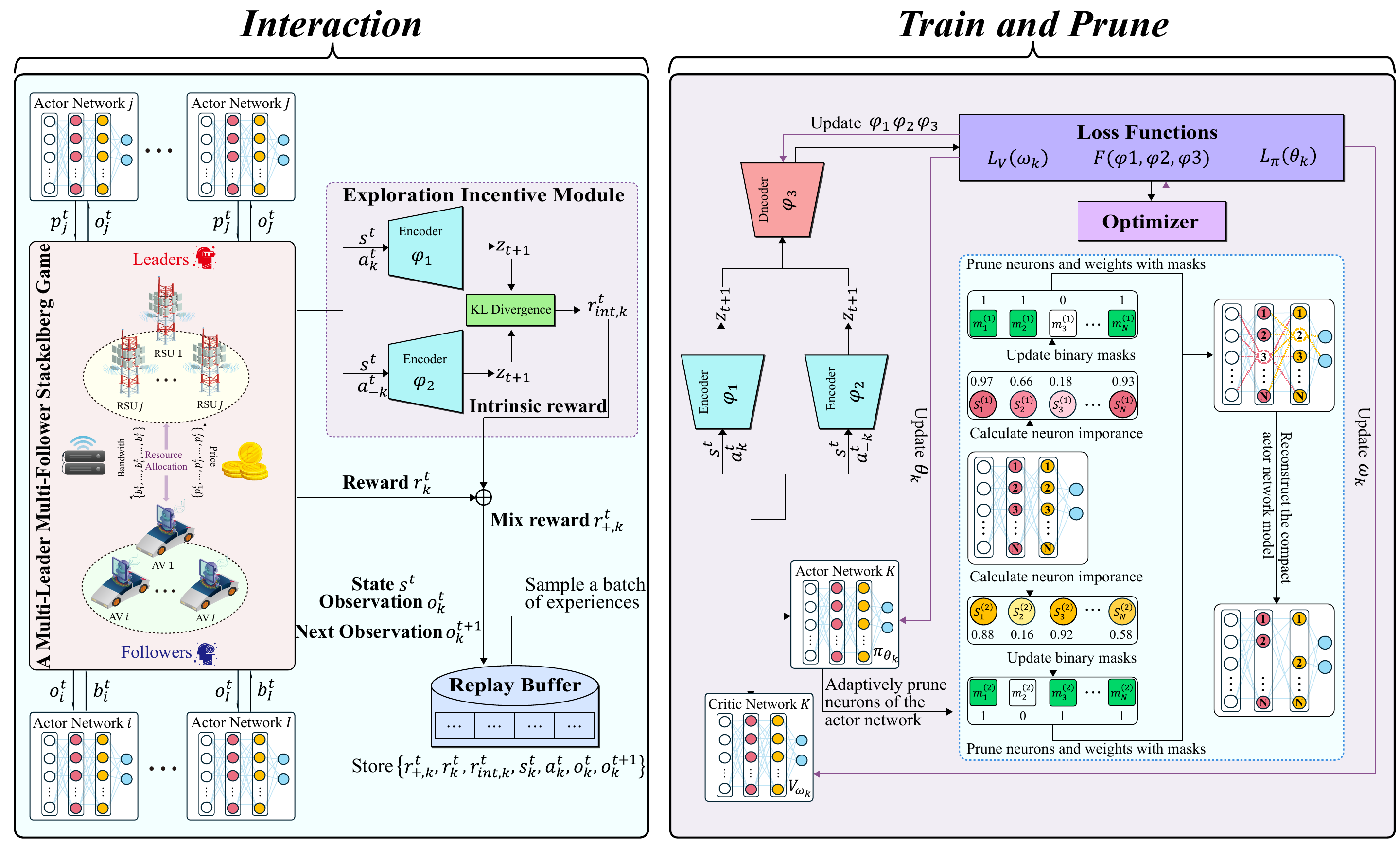}
    \caption{TinyIEI-MAPPO algorithm’s Architecture for the VEAAT migration. }
    \label{fig:algorirhm}
\end{figure*}
 
\section{A Tiny Multi-Agent Reinforcement Learning Algorithm with Self-Adaptive Dynamic Structured Pruning}\label{tiny algorithm}

In MLMF Stackelberg games between RSUs and AVs, complex data transmission and decision-making face challenges like privacy, incomplete info, and environmental dynamics. Traditional heuristic algorithms are unsuitable, while DRL shows promise but has drawbacks: in high-dimensional solution spaces, it lacks proper exploration mechanisms, leading to low sample efficiency and high computing resource consumption \cite{loftin2021strategically}; as model size grows, neural networks have redundant components, increasing computational burden and risking overfitting \cite{song2023comprehensive}. To address these issues and enhance algorithm efficiency and performance, this section proposes TinyMA-IEI-PPO, a tiny multi-agent reinforcement learning algorithm with dynamic adaptive structural pruning based on individual exploration incentive, first modeling the game as a Partially Observable Markov Decision Process (POMDP), then introducing individual exploration incentive during training and adaptively adjusting pruning thresholds in the pruning phase.

\subsection{Deep Reinforcement Learning Preliminaries for Stackelberg Game}
We first represent the MLMF Stackelberg game as a multi-agent POMDP. Specifically, let a POMDP be represented by the tuple $\langle\mathcal{S},\mathcal{O},\mathcal{A},\mathcal{T},\mathcal{R}\rangle$, where $\mathcal{S},\mathcal{O},\mathcal{A},\mathcal{T}$, and $\mathcal{R}$ respectively represent the state space, observation space, action space, set of state-transition probability functions, and reward function. We describe the detailed definitions of each term as follows.

\begin{enumerate}
    \item \textbf{State Space:}  We denote $\mathcal{S}\triangleq\{s^1,\ldots,s^t\,\ldots\}$ as the global observation space. At each time step $t$ within the time series $\mathcal{T}=\{0,\ldots,t,\ldots,T\},$ the state space is defined by $s^t\triangleq\{\boldsymbol{P}^t,\boldsymbol{B}^t,\boldsymbol{\Lambda}^t,\boldsymbol{\mu}^t,\boldsymbol{R}_L^t,\boldsymbol{R}_F^t\}.$ Here, $\boldsymbol{P}^t$ is the pricing strategy of all RSUs, $\boldsymbol{B}^t$ is the bandwidth-demand strategy of the AVs, reflecting its network resource requirements. The task arrival rate $\boldsymbol{\Lambda}^t$ and the task processing rate $\boldsymbol{\mu}^t$ represent the dynamic situation of system tasks, where $\boldsymbol{\Lambda}^t=\{\lambda^t_{1},\ldots,\lambda^t_{m},\ldots,\lambda^t_{j}\}\mathrm{~and~}\boldsymbol{\mu}^t=\{\mu^t_{1},\ldots,\mu^t_{m},\ldots,\mu^t_{j}\}.$ 
     \item \textbf{Partially Observable Space:} Due to privacy protection, agents are unable to obtain the complete state of the environment and can only make decisions based on local observations. In the initial stage of training when $t<L,\boldsymbol{P}^{t-L}\mathrm{~and~}\boldsymbol{B}^{t-L}$ are randomly generated. At the beginning of each training time step $t$, in stage I, RSU $j$ determines its pricing strategy $p_j$ based on the past pricing strategies of RSUs, the bandwidth-demand strategies of AVs in the past $L$ rounds, the current task arrival rate $\lambda_{tj}$ and the task processing rate $\mu^t_{j}$. Its observation space is $o^t_{j}\triangleq\{\boldsymbol{B}^{t-L},\boldsymbol{P}^{t-L},\ldots,\boldsymbol{B}^{t-1},\boldsymbol{P}^{t-1},\lambda^t_{j},\mu^t_{j}\}.$ In stage II, AV $i$ determines its bandwidth-purchase strategy $b_{ij}$  by referring to the historical pricing strategies of RSUs and the historical bandwidth-demand strategies of other AVs. Its observation space is $o^t_{i}\triangleq\{\boldsymbol{B}^{t-L}_{-j},\boldsymbol{P}^{t-L},\ldots,\boldsymbol{B}^{t-1}_{-j},\boldsymbol{P}^{t-1}\}.$
    \item \textbf{Action Space:} We denote $\mathcal{A}^k\triangleq\{a^k\}$ as the action space of agent $k$. At each time step $t$, for RSU $j$, considering the migration cost \(c_j\) and the upper-bound price \(p_{max}\) for the pricing action, the action space is defined as \(a^t_j=p^t_{j}\in[c_j,p_{max}]\). AV \(i\) determines vector \(a^t_i=B^t_{i}=\{b_{ij}\}_{j \in\mathcal{R}}\), which represents the bandwidth demand for each RSU \(j\), and the value range is \([0,+\infty)\). The decision-making processes of both of them rely on the information encapsulated in the partially observable space.
    \item \textbf{Reward Functioin:} The internal reward function of agent $k$, aligned with Eq.(\ref{individual exploration incentives}), is denoted as $r_{int,k}^t$. The hybrid reward function of agent $k$, aligned with Eq.(\ref{mix rew}), is denoted as $r_{+,k}^t$. The hybrid rewards of all agents are represented by $\boldsymbol{R}_+^t$. The rewards of all RSUs and AVs are $\boldsymbol{R}_L^t$ and $\boldsymbol{R}_F^t$ respectively, where $\boldsymbol{R}_L^t=\{r^t_{1},\ldots,r^t_{m},\ldots,r^t_{j}\}$ and $\boldsymbol{R}_F^t=\{\tau^t_{1},\ldots,\tau^t_{n},\ldots,\tau^t_{i}\}$. The reward functions of RSU \(j\) and AV \(i\) are defined based on the utility functions Eq.(\ref{Ulity of L}) and Eq.(\ref{Ulity of F}) from our Stackelberg game. At time step \(t\), the reward of RSU \(j\) is $r^t_j = U^L_j(p^t_j,\boldsymbol{P}^t_{-j},\boldsymbol{B}^t)$, and the reward of AV \(i\) is $\tau^t_i = U^F_i(\boldsymbol{B}^t_i,\boldsymbol{B}^t_{-i},\boldsymbol{P}^t)$.
\end{enumerate}

We employ the Multi-Agent Proximal Policy Optimization (MAPPO) algorithm and make improvements based on it. The policy \(\pi(o^t,\theta)\) is parameterized by an actor network with the weight parameter \(\theta\), while the state value  \(V(o^t,\omega)\) is parameterized by a critic network with the weight parameter \(\omega\). For the \(k\)-th agent, the loss function of the critic network is obtained by minimizing the expected value of the square of the TD (Temporal Difference) error \(\delta\), which can be represented as
\begin{equation}\min_{\boldsymbol{\omega}_{k}}L_{V}(\boldsymbol{\omega}_{k})=\min_{\boldsymbol{\omega}_{k}}\mathbb{E}\left[\left(r_{k}^t+\gamma_kV({\boldsymbol{\omega}_{k}},\boldsymbol{o}_{k}^t)-V({\boldsymbol{\omega}_{k}},\boldsymbol{o}_{k}^t)\right)^{2}].\right.
\label{critic_function}
\end{equation} 

The objective of policy iteration is defined as
\begin{equation}
    \begin{split}
        \max_{\boldsymbol{\theta}_{k}}L_{\pi}(\boldsymbol{\theta}_{k})=\max_{\boldsymbol{\theta}_{k}}&\mathbb{E}\left[\min\left(f^t_{k}(\boldsymbol{\theta}_{k})\hat{A}_{\pi_{\boldsymbol{\theta}_k}}(\boldsymbol{o}_{k},\boldsymbol{a}_{k}),\right.\right.\\
        &\left.\left.g_{clip}\left(f^t_{k}(\boldsymbol{\theta}_{k})\right)\hat{A}_{\pi_{\boldsymbol{\theta}_{k}}}(\boldsymbol{o}_{k},\boldsymbol{a}_{k})\right)\right],
        \label{actor_loss_functioin}
    \end{split}
\end{equation}
where $f^t_k(\theta_k)=\frac{\pi_{\theta_k}(a_t|o_t)}{\pi_{\theta^{oid}_k}(a_t|o_t)}$. $ f^t_k(\theta_k)$ is an importance ratio function, which measures the difference between the current policy $\pi_{\theta_k}(a_t|o_t)$ and the old police $\pi_{\theta^{oid}_k}(a_t|o_t)$.  \(\hat{A}_{\pi_{\boldsymbol{\theta}_k}}(\boldsymbol{o}_{k},\boldsymbol{a}_{k})\) is an estimator of the advantage function, which is calculated as 
\begin{equation}\hat{A}_{\pi_{\boldsymbol{\theta}_k}}\left(o^t_k,a^t_k\right)=-V_{\pi_{\boldsymbol{\theta}_k}}\left(o^t_k\right)+\sum_{l=0}^\infty(\gamma_k)^lr(t+l).\end{equation}

The clipping function is defined as \begin{equation}g_{clip}(f^t_{k}(\boldsymbol{\theta}_{k}))=
\begin{cases}
1-\epsilon,f^t_{k}(\boldsymbol{\theta}_{k})<1-\epsilon, \\
f^t_{k}(\boldsymbol{\theta}_{k}),1-\epsilon\leq f^t_{k}(\boldsymbol{\theta}_{k})\leq1+\epsilon, & 
 \\
1+\epsilon,f^t_{k}(\boldsymbol{\theta}_{k})>1+\epsilon,
\end{cases}\end{equation}
where $\epsilon$ is an adjustable hyperparameter.
Its main purpose is to constrain the importance ratio \(f^t_{k}(\boldsymbol{\theta}_{k})\). When \(f^t_{k}(\boldsymbol{\theta}_{k})\) exceeds a certain range, \(g_{clip}\left(f^t_{k}(\boldsymbol{\theta}_{k})\right)\) will adjust its value to fall within an appropriate interval. This adjustment mechanism is vital as it prevents large fluctuations during policy updates, ensuring the algorithm's stability.

\subsection{Individual Exploration incentives as Intrinsic Incentives}
In multi-agent reinforcement learning, the exploration strategy plays a crucial role in enabling agents to discover optimal policies. In this section, we will introduce an agent-level intrinsic exploration module solely for training, which will be removed after training to avoid latency impacts and how to characterize and estimate the individual exploration incentives and let them serve as intrinsic incentives within the MAPPO framework.

\subsubsection{Bayesian Surprise to Characterize Individual Exploration Incentives}
We focus on evaluating the individual exploration incentives of a specific action \(a^t_{k}\) performed by agent \(k\), denoted as \(r^t_{k, int}\). The objective is to assess and prompt agents to take actions that significantly affect global latent state transitions, rather than those with the highest value. Based on prior work \cite{li2024individual}, we use the Bayesian surprise rate to measure the difference between actual and counterfactual latent-state distributions from agent $k$'s perspective. We represent the individual exploration incentives \(r^t_{k, int}\) as the mutual information between the latent variable \(z^{t + 1}\) and the action \(a^t_{k}\), which is expressed as
\begin{equation}
\begin{split}
  r^t_{k,int}&=I(z^{t+1};a^t_{k}|s^t,a^t_{-k})\\&=D_{KL}\left[p(z^{t+1}|s^t,a^t)\parallel p(z^{t+1}|s^t,a^t_{-k})\right].  
  \label{individual exploration incentives}
\end{split}\end{equation}
 where $s^t$ is the global observation in time step $t$.

\subsubsection{Conditional Variational Autoencoder (CVAE) to Estimate the Bayesian Surprise}
To robustly estimate individual exploration incentives, we leverage the CVAE to resolve the problem of latent space misalignment and precisely identify the latent space of \(z_t\) for reconstructing the original state space by utilizing the CVAE. The overall architecture and training process of the CVAE is illustrated in Fig.\ref{fig:algorirhm}.
The training objective of this module is to maximize the variational lower bound of the conditional log-likelihood, which is as follows:
\begin{equation}
\begin{aligned}
\mathcal{F}&(\varphi_1,\varphi_2,\varphi_3) = \\&-D_{\mathrm{KL}}\left[q_{\varphi_1}(z^{t + 1}|s^{t},\boldsymbol{a}^{t},s^{t + 1})\|p_{\varphi_1}(z^{t + 1}|s^{t},\boldsymbol{a}^{t})\right] \\
& -D_{\mathrm{KL}}\left[q_{\varphi_2}(z^{t + 1}|s^{t},\boldsymbol{a}^{t}_{-k},s^{t + 1})\parallel p_{\varphi_2}(z^{t + 1}|s^{t},\boldsymbol{a}^{t}_{-k})\right] \\
& +\mathbb{E}_{z\sim q_{\varphi_1}}\left[\log p_{\varphi_3}(s^{t + 1}|z)\right] +\mathbb{E}_{z\sim q_{\varphi_2}}\left[\log p_{\varphi_3}(s^{t + 1}|z)\right].
\end{aligned}
\end{equation}

\subsubsection{Harnessing Individual Exploration Incentives to Improve PPO's Loss Function}
Subsequently, we introduce these individual exploration Incentives as intrinsic motivation and combine them with the external rewards to form a hybrid reward for agent training, as shown in Eq.(\ref{mix rew}),
\begin{equation}r^t_{+,k}(s^t,a^t)=r^t_k(s^t,a^t)+c_1 r^t_{k,int},\label{mix rew}
\end{equation}
where $c_1$ is a hyperparameter to balance intrinsic incentives and external environment rewards in training, since the scales are different. Correspondingly, to continuously guide agents to explore more purposefully during training, we have also made the following improvements to the objective function of PPO:
\begin{equation}L_{ppo}=\mathbb{E}
\begin{bmatrix}
L_{\pi}(\boldsymbol{\theta}_{k})-c_2L_{V}(\boldsymbol{\omega}_{k})+c_3(\mathbb{E}_{\pi_k}[r_{\mathrm{int}}^k]+H(\pi_{\theta_k}|\tau))
\end{bmatrix},\end{equation}
where  $H(\pi_{\theta_k}|\tau)=-\beta\mathbb{E}_{\pi_{\theta_k}(\cdot|\tau)}\ln\pi_{\theta_k}(\cdot|\tau;\theta_k)$ is the policy entropy and $\beta$ is a hyper-parameter to control the regularization weight for entropy maximization. Both \(c_2\) and \(c_3\) are hyperparameters that control the weight of each term in the PPO loss function. In particular, to distinguish between the exploration and exploitation stages and ensure the convergence of training,\(c_3\) is designed as an annealing weight parameter that gradually decays as the training progresses, which can be expressed as follows:
\begin{equation}c_3=\frac{e}{1 + e^{\alpha(N - N_0)}}.\end{equation}
Here, \(\alpha\) is a hyperparameter controlling the annealing rate, \(N\) represents the number of training steps and \(N_0\) is an offset parameter denoting the step number at which the annealing process starts to decline.

\begin{algorithm}
\caption{TinyMA-IEI-PPO-based Solution for MLMF Stackelberg Game}
\label{alg:tinyMAP_IEI_PPO} 
\begin{algorithmic}[1]
\Require A DRL training environment $\mathcal{E}$; A lightweight Tiny and Compact MADRL model $\mathcal{M} (\theta_k, w_k)$ of agent $k$ with exploration incentive module parameters $\varphi_1, \varphi_2, \varphi_3$; Maximum episodes $E$, update time $L$, maximum time step $T$.
\Ensure A trained tiny model with optimized parameters for the MLMF Stackelberg Game.
\For{agent $k \in \mathcal{R}\cup\mathcal{V}$}
    \State Initialize $\pi_{\theta_k}, V_{\omega_k}$.
\EndFor
\While{episode $e \leq E$}
    \State Reset Stackelberg game environment, get state $S_0$ and reply buffer $\mathcal{D}_k$.
    \For{time step $t \in 1, \ldots, T$}
        \State Input $o^t_j$ into $j$-th RSU's actor policy $\pi_{\theta_j}$ and determine the price strategy $p^t_j$.
        \State Input $o^t_i$ into $i$-th AV's actor policy $\pi_{\theta_i}$ and determine the bandwidth demand strategy $b^t_i$.
        \State Calculate utility function for AV $i$ and RSU $j$ through Eq.(\ref{U_F}) and Eq.(\ref{U_L}).
        \State Calculate individual exploration incentive $r^t_{int,k}$ and the hybrid reward $r^t_{+,k}$ by Eq.(\ref{individual exploration incentives}) and Eq.(\ref{mix rew}).
        \State Calculate the current neuron importance $\phi^{(L)}_n$ by Eq.(\ref{current importance}).
        \State Update $S_t$ to $S_{t + 1}$.
        \State $\mathcal{D}_k = \mathcal{D}_k \cup \{o_k, a^t_k, R^t_k, o^t_{k + 1}, r^t_{int,k},r^t_{+,k}\}$.
        \If{$t$ $\bmod$ train-interval $== 0$}
            \State Update $\varphi_1, \varphi_2, \varphi_3$ using the procedure in Algorithm \ref{alg:train_exploration_incentive} with inputs $\varphi_1, \varphi_2, \varphi_3, \mathcal{D}$.
            \State Update $\omega_k, \theta_k$ using the procedure in Algorithm \ref{alg:train_policy_prune} with inputs $\omega_k, \theta_k, \mathcal{D}$.
        \EndIf
    \EndFor
\EndWhile
\end{algorithmic}
\end{algorithm}

\begin{algorithm}[t]
\caption{Train Exploration Incentive Module: Training procedure of Exploration Incentive Module}
\label{alg:train_exploration_incentive} 
\begin{algorithmic}[1]
\Require Exploration Incentive Module parameters $\varphi_1, \varphi_2, \varphi_3$, replay buffer $\mathcal{D}$.
\Ensure Optimized Exploration Incentive Module parameters.
\State Sample batch $\sim \mathcal{D}$.
\State Update $\varphi_1 \leftarrow \varphi_1 +$ learning rate $\cdot \nabla \mathcal{F}_{\varphi_1}(\varphi_1, \varphi_2, \varphi_3)$.
\State Update $\varphi_2 \leftarrow \varphi_2 +$ learning rate $\cdot \nabla \mathcal{F}_{\varphi_2}(\varphi_1, \varphi_2, \varphi_3)$.
\State Update $\varphi_3 \leftarrow \varphi_3 +$ learning rate $\cdot \nabla \mathcal{F}_{\varphi_3}(\varphi_1, \varphi_2, \varphi_3)$.
\end{algorithmic}
\end{algorithm}

\begin{algorithm}[t]
\caption{Train Policy and Prune: Self-Adaptive Dynamic Structural Pruning for MADRL Network}
\label{alg:train_policy_prune} 
\begin{algorithmic}[1]
\Require DRL Network parameters $\theta_k, \omega_k$.
\Ensure A Tiny and Compact DRL model $(\theta_K, \omega_k)^{(L)}$.
\State Calculate the loss $L_{\pi}(\theta_k), L_V(\omega_k)$ by Eq. (\ref{utl_actor_function}) and (\ref{critic_function}).
\State Calculate the Neuron-Importance Metric based on Time-window Decay $S^{t,(l)}_n$ by Eq. (\ref{importance}).
\State Update the actor network parameter $\theta^{(L)}_k$ by Eq.(\ref{update actor}).
\State Update the critic network parameter $\omega^{(L)}_k$ by Eq.(\ref{update critic}).
\State Calculate the adaptive pruning threshold $\psi$ by Eq.(\ref{threshold}).
\State Updating the mask $m^{t,(l)}_n$ by Eq.(\ref{mask}).

\For{ each neuron $\mathcal{N}$ in the actor network}  network 
    \If{$ S^{t,(l)}_n < \varphi$}
        \State Remove $\mathcal{N}^{(l)}$ and parameters connected to the removed neuron.
    \EndIf
\EndFor
\State Reconstruct a Tiny and Compact DRL model $(\theta_k, \omega_k)^{(L)}$.
\end{algorithmic}
\end{algorithm}
\subsection{The Approach of Adaptive Dynamic Structure Pruning Based on Individual Exploration Incentives}
The adaptive dynamic structural pruning algorithm based on individual exploration incentives proposed in this paper the algorithm encompasses three key steps: (i) establishing an accurate neuron importance metric, (ii) adaptively determining the pruning threshold according to the individual exploration incentives, and (iii) updating the binary mask for pruning.
\subsubsection{Neuron-Importance Metric based on Time-window Decay}
In terms of network architecture, both the actor network and the critic network have a fully connected network structure. For a given actor network with L layers, we use \(h\) to denote the hidden layers, excluding the input and output layers. We represent the weights in the \(l\)-th fully-connected layer as \(\boldsymbol{\theta}^{(l)}\). At time step \(t\), the neuron importance \(\Omega^{t,(l)} _n\) of the \(n\)-th neuron in the \(l\)-th layer can be expressed as follows \cite{su2024compressing}:

\begin{equation}
\Omega_{n}^{t,(l)}t=\sum_{n}\left(\theta_{m,n}^{t,(l)}\right)^{2}\cdot\sum_{o}\left(\theta_{o,m}^{t,(l+1)}\right)^{2}.
\label{current importance}
\end{equation} 
Removing such neurons can compromise the network’s capacity to learn effective strategies.
To address this issue, we propose the Time-window Dynamic Decay Neuron-Importance Metric. By integrating a time window and a forgetting function, this metric effectively reduces noise interference during early training. It can more accurately capture the important changes of neurons across different time steps, as detailed below:

\begin{equation}S _n^{t,(l)}=\sum_{\tau=t-t_W}^t\gamma _n^{(w-\tau)}\Omega _n^{\tau,(l)}\cdot m _n^{t,(l)},
\label{importance}
\end{equation}

where \(t_w\) is the starting time step of the time window, \(w\) is the width of the time window, \(\gamma _n\) is the decay factor, and \(m _n^{t,(l)}\) is used for the pruning status of $\mathcal{N}$-th neuron in the \(l\)-th layer.

\subsubsection{Dynamic pruning threshold adaptive to the individual exploration incentives}
To enable the model to better adapt to structural changes, we adopt a more refined pruning approach where the model sparsity gradually increases with the number of iterations \cite{livne2020pops}. Moreover, to encourage the model to conduct more effective exploration and preserve its exploration ability to search for the equilibrium solution of the Stackelberg game in the early stage, we evaluate the exploration incentives of agents and further dynamically fine-tune the pruning threshold on the basis of the original dynamic pruning strategy. Therefore, the definition of the dynamic pruning threshold adaptive to the exploration incentives of individuals is as follows:
\begin{equation}
\psi=\sum _n\sum_lS _n^{(l)}\cdot p_t,
\label{threshold}
\end{equation}

\begin{equation}
\begin{split}
    p_{t} &= \min\left\{
        \max\left(
            p_{t1} \cdot \left(1 + \phi r_{k,int}^{t^{\prime}}\right),
            \right.
            \right.\\
        &\quad\left.
            \left.
            p_{t2}\cdot\left(1 + \phi r_{k,int}^{t-1^{\prime}}\right)
        \right),
        p_{t1}
    \right\},
\end{split}
\end{equation}
where \(\psi\) is the pruning threshold, \(\phi\) is a hyperparameter used to control the sensitivity of the pruning threshold to the individual exploration degree, and \(p_t\) is the pruning rate after the adaptive adjustment of the original pruning strategy. The original progressive pruning strategies \(p_{t1}\) and \(p_{t2}\) are given by the following equations:

\begin{equation}p_{t1}=p_f+(p_i-p_f)\left(1-\frac{t-t_0}{N\Delta t}\right)^4,\end{equation}
\begin{equation}p_{t2}=p_f+(p_i-p_f)\left(1-\frac{t-t_0}{N\Delta t}\right)^2,\end{equation}
where $p_i$ is the initial sparsity, $p_f$ is the target sparsity, $t_0$ is the starting epoch of gradual pruning,$ N $is the total pruning steps, and $\Delta$ is the pruning frequency.

In order to more convenient for the algorithm to perceive and effectively utilize the individual exploration degree \(r_{k, int}^t\) to dynamically adjust the pruning rate and the pruning threshold,  we transform (\ref{individual exploration incentives}) by using the Jensen-Shannon (JS) divergence:
\begin{equation}
\begin{aligned}
r_{k,int}^{t^{\prime}} &= D_{JS}(p\parallel q) \\
&= \frac{1}{2}D_{KL}\left(p\parallel\frac{p + q}{2}\right) + \frac{1}{2}D_{KL}\left(q\parallel\frac{p + q}{2}\right), \\
\end{aligned}
\end{equation}
where $p = p(z^{t + 1}|s^{t},a^{t}),  q = p\left(z^{t + 1}|s^{t},a^{t}_{-k}\right).$The JS divergence is symmetric, and its value range is between 0 and 1.

\subsubsection{Update the pruning binary mask}
For a given actor network that contains $L$ layers, we denote the hidden layers between the input and output layers by $h$. Since the output of one layer is the input of the next, the output of the $l$-th layer can be expressed as
\begin{equation}h^{(l)}=\sigma^{(l)}\left(\theta^{(l)}h^{(l-1)}\odot m^{(l)}\right),\end{equation}
where $\sigma^{(l)}$ represents the nonlinear response of the output layer. At the beginning of the training, all elements of the binary mask $m$ are initialized to 1, indicating that the corresponding neurons should be retained. The symbol $\odot$ indicates the element-wise multiplication of two matrices.

We integrate the binary mask with the actor-network, and the loss function thereof is restated as follows:
\begin{equation}\begin{aligned}
 & \textbf{\textit{P3:}}\max_{\boldsymbol{\theta}}L_{\pi}(\boldsymbol{\theta}) \\
 & \mathrm{s.t.}\sum_{l=1}^{L-1}\|m^{(l)}\|_{0}\leq C.
 \label{rewrite_1}
\end{aligned}\end{equation}

The $\|\cdot\|_0$ is the zero-norm, which represents the number of non-zero elements. $C$ is a hyperparameter that governs the quantity of pruned neurons.
After calculating the value of the dynamic pruning threshold, we sort the neurons in ascending order of their importance. Neurons that rank below the threshold are removed, while those that rank above the threshold are retained. Therefore, the mask is updated as follows:
\begin{equation}m_n^{(l)}=
\begin{cases}
 0, & \mathrm{if~abs}\left(m_n^{(l)}\theta_n^{(l)}\right)<\psi, \\
1, & \mathrm{if~abs}\left(m_n^{(l)}\theta_n^{(l)}\right)\geq\psi,
\label{mask}
\end{cases}\end{equation}
where, $abs(\cdot)$ represents the absolute value. Eq.(\ref{rewrite_1}) can be transformed into the following Lagrangian multiplier-based form. Moreover, by integrating the binary mask with Eq.(\ref{importance}), we construct the Neuron-Importance Group Sparse Regularizer \cite{su2024compressing} \cite{kang2024tiny}. Consequently, we can integrate the update of the binary mask with the update of the actor network. During the training phase, the binary mask is updated simultaneously, and the neurons whose importance is lower than the pruning threshold are removed. Thus, Eq.(\ref{rewrite_1}) can be rewritten as follows:
\begin{equation}
\begin{aligned}
&\max_{\boldsymbol{\theta} }\mathbb{E}[\min(f^t (\boldsymbol{\theta} )\hat{A}_{\pi_{\boldsymbol{\theta}}}(\boldsymbol{o} ,\boldsymbol{a} ),g_{clip}\left(f^t (\boldsymbol{\theta})\right)\hat{A}_{\pi_{\boldsymbol{\theta} }}(\boldsymbol{o} ,\boldsymbol{a} ))]\\
&\quad \quad \quad \quad\quad\quad\quad +c_2(\mathbb{E}_{\pi_k}[r_{\mathrm{int}}^k]+H(\pi_{\theta_k}|\tau)-\lambda\sum _n\sum_lS _n^{(l)}.
\label{utl_actor_function}
\end{aligned}
\end{equation}

Therefore, the actor-network parameters $\theta$ and the critic-network parameters $\omega$ can be iteratively updated via stochastic gradient ascent, as expressed in the following update equations:

\begin{equation}\theta^{(l)}\leftarrow\theta^{(l)}-l_{actor}\frac{\partial L_{\pi}(\boldsymbol{\theta})}{\partial(h^{(l)}\odot m^{(l)})}\cdot\frac{\partial(h^{(l)}\odot m^{(l)})}{\partial\theta^{(l)}},
\label{update actor}
\end{equation}

\begin{equation}w^{(l)}\leftarrow w^{(l)}-l_{critic}\frac{\partial L_{V}({\omega})}{\partial w^{(l)}},
\label{update critic}
\end{equation}
where $l_{actor}$ and $l_{critic}$ are learning rates of the gradient descent. Building on the aforementioned analysis, the overall TinyMA-IEI-PPO algorithm is presented in \textbf{Algorithm \ref{alg:tinyMAP_IEI_PPO}}.

\section{Numerical Results}\label{Results}
\begin{figure}[H] 
    \centering
    \includegraphics[width=0.9\linewidth]{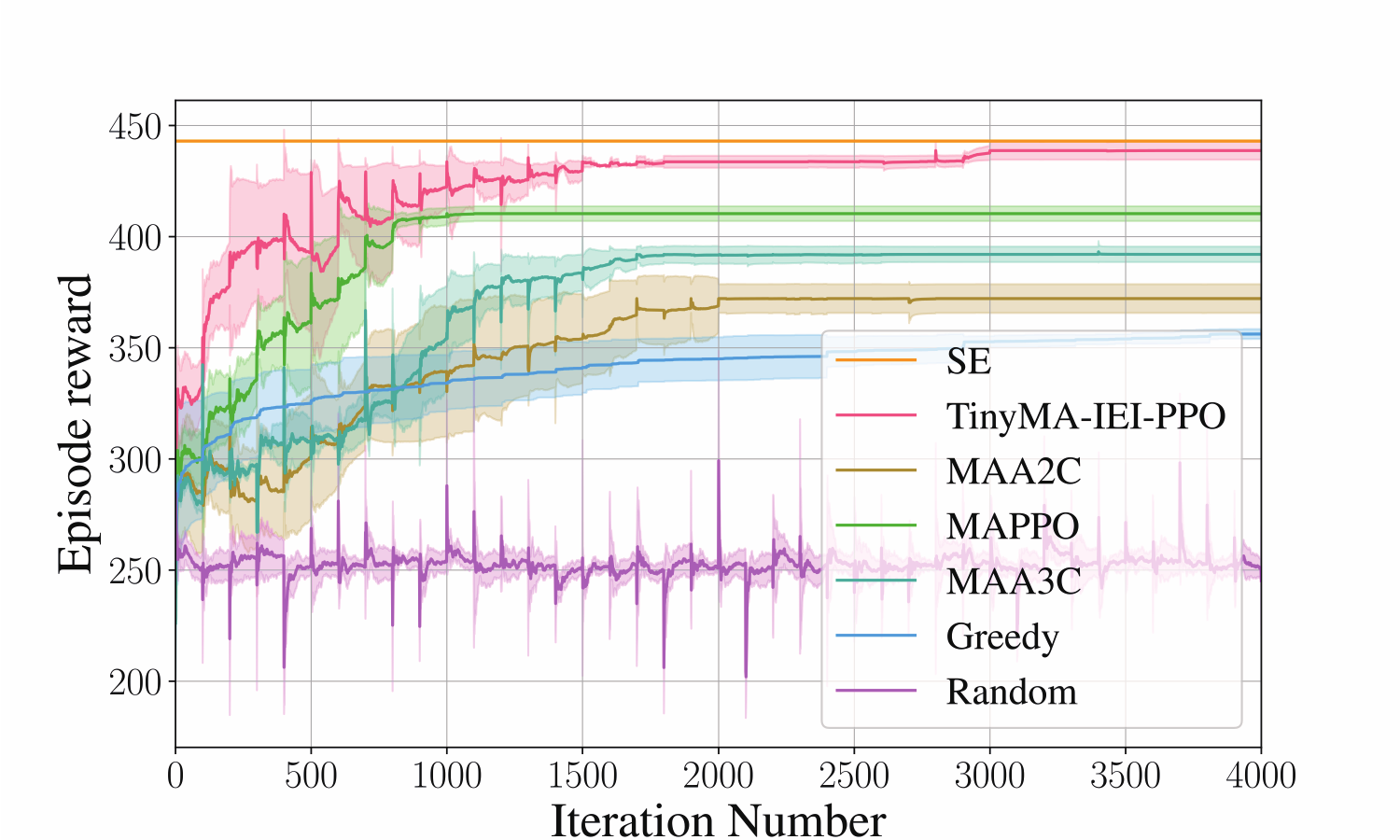}
    \caption{Comparison of episode reward curves of TinyMA-IMI-PPO and baselines for the MFML Stackelberg Game.}
    \label{TinyMA_IEI_PPO_baselines}
\end{figure}
\begin{figure*}
    \centering
    \begin{minipage}[t]{0.45\textwidth}
        \centering
        \includegraphics[width=\textwidth]{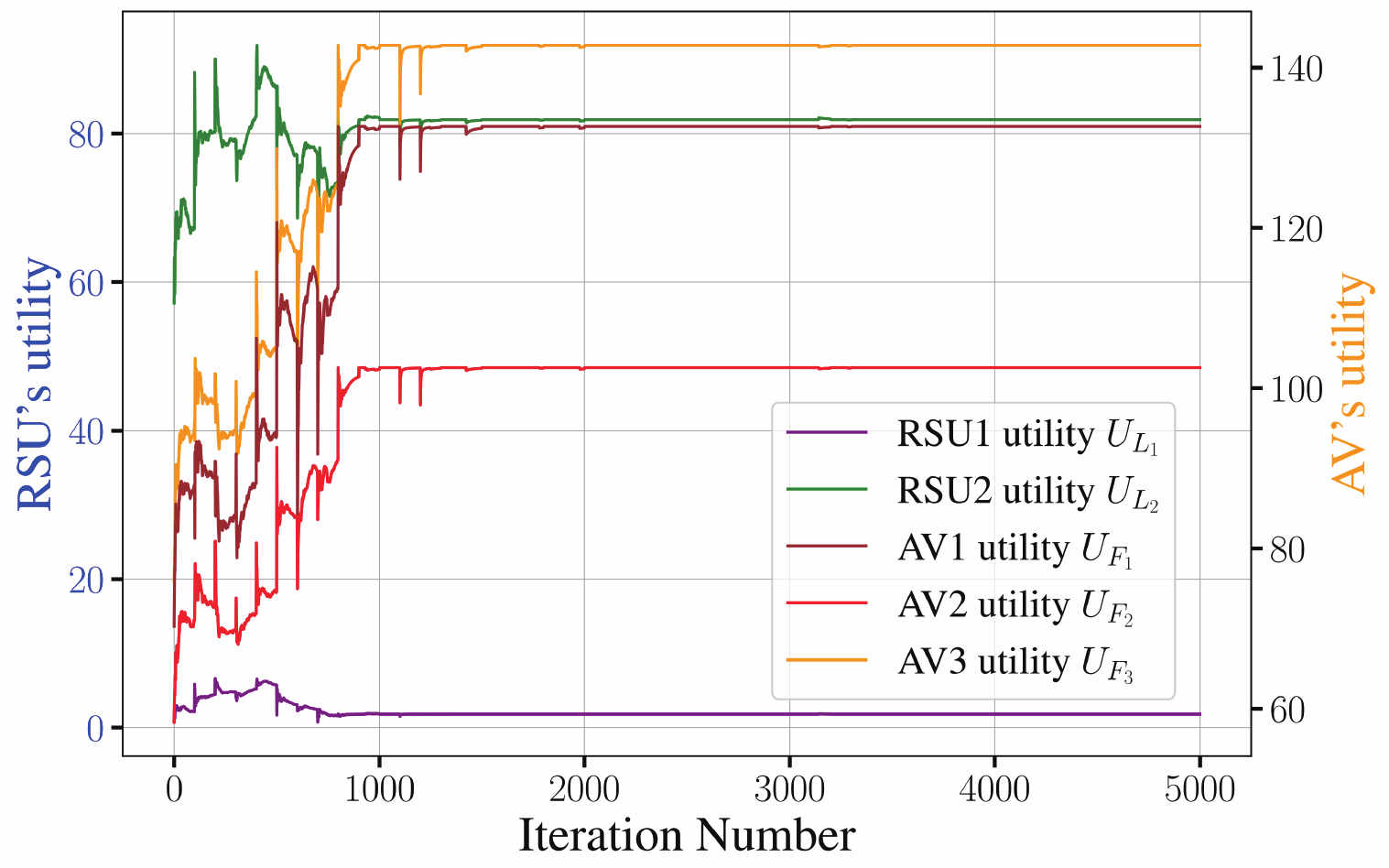}
        \caption{Utilities of AVs and RSUs}
        \label{rew_plot}
    \end{minipage}
    \hfill
    \begin{minipage}[t]{0.45\textwidth}
        \centering
        \includegraphics[width=\textwidth]{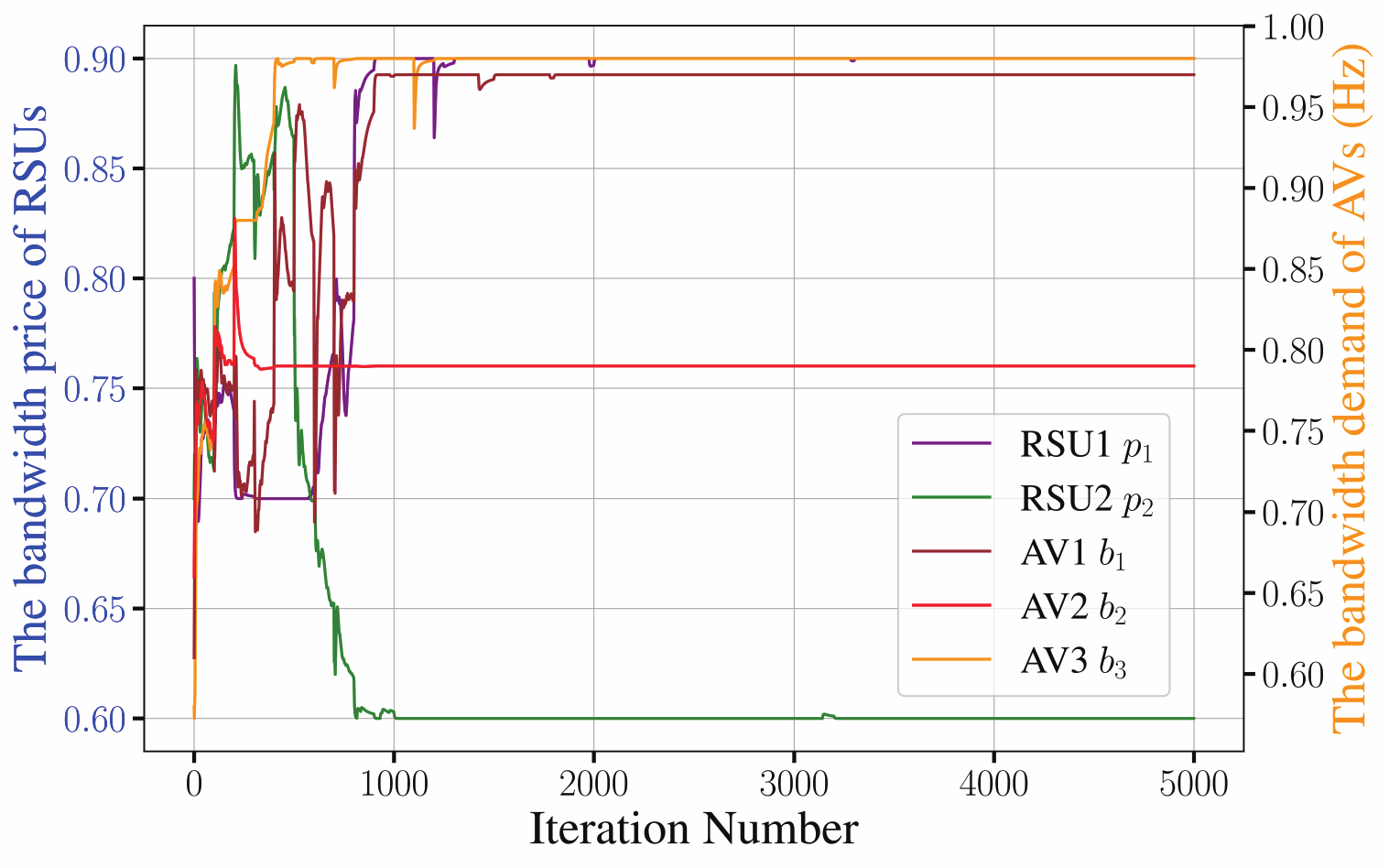}
        \caption{The pricing strategies of RSUs and the bandwidth demands of AVs}
        \label{act_plot}
    \end{minipage}
    \label{Reward-Num-change}
\end{figure*}

\begin{figure*}
    \centering
    \begin{minipage}[t]{0.23\textwidth}
        \centering
        \includegraphics[width=\textwidth]{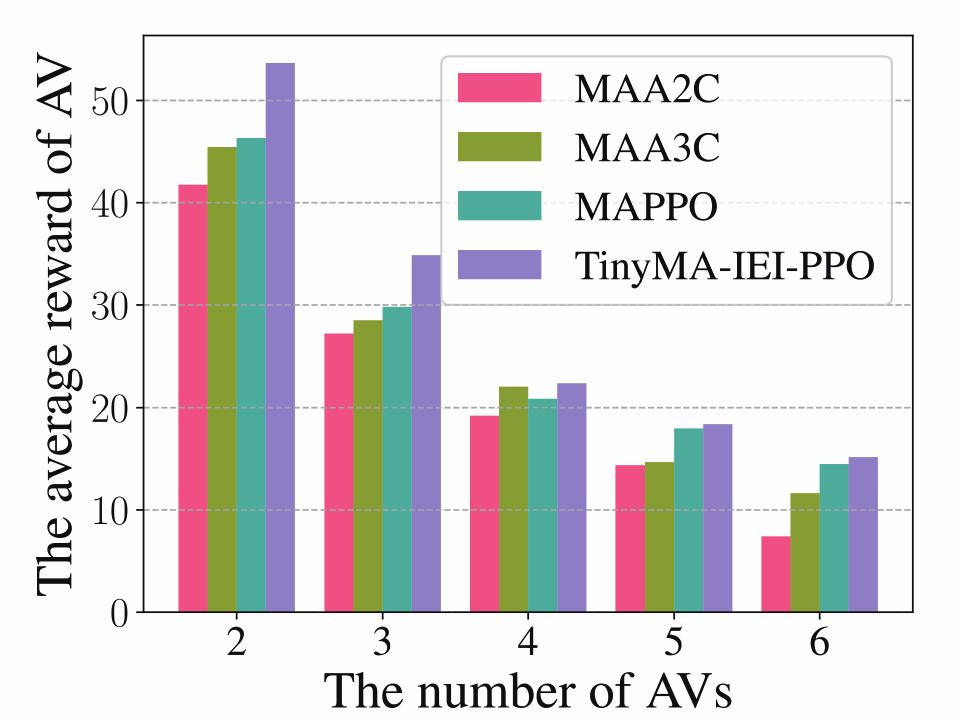}
        \caption{The average reward of AVs for different numbers of AVs underdifferent algorithms.}
        \label{AV_reward_AV2}
    \end{minipage}
    \hfill
    \begin{minipage}[t]{0.23\textwidth}
        \centering
        \includegraphics[width=\textwidth]{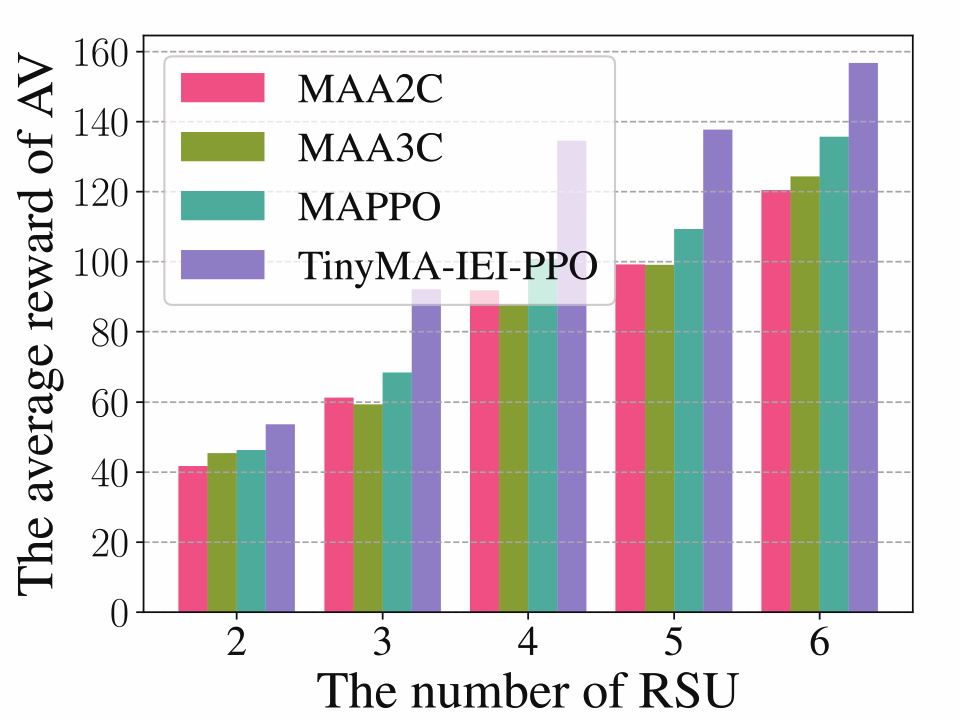}
        \caption{The average reward of AVs for different numbers of RSUs underdifferent algorithms.}
        \label{AV_reward_RSU2}
    \end{minipage}
    \hfill
    \begin{minipage}[t]{0.23\textwidth}
        \centering
        \includegraphics[width=\textwidth]{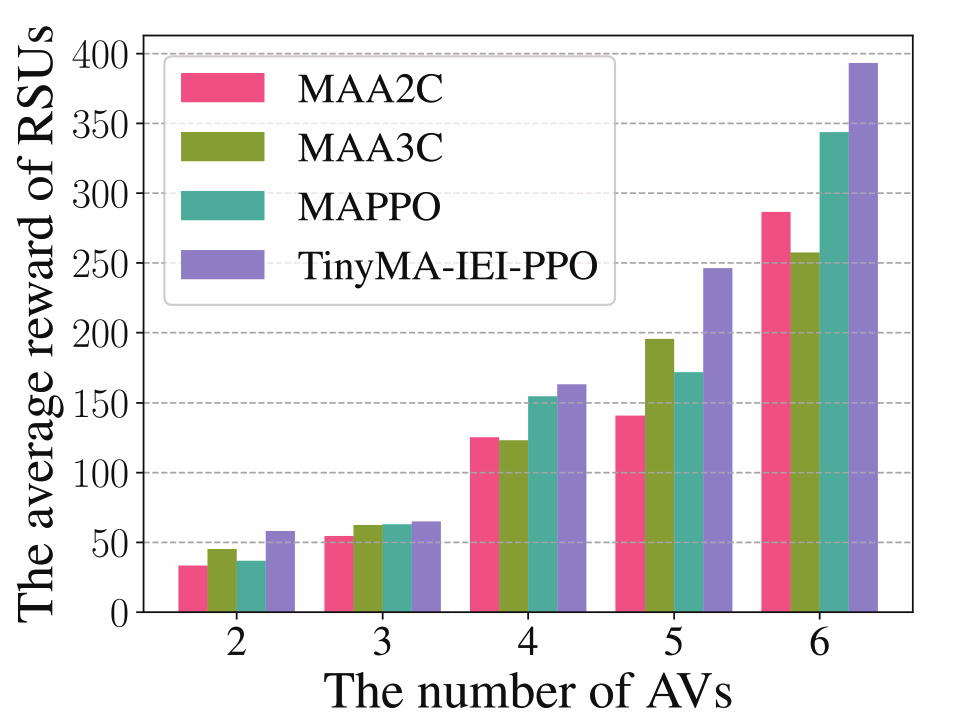}
        \caption{The average reward of RSUs for different numbers of AVs under different algorithms.}
        \label{RSU_reward_AV2}
    \end{minipage}
    \hfill
    \begin{minipage}[t]{0.23\textwidth}
        \centering
        \includegraphics[width=\textwidth]{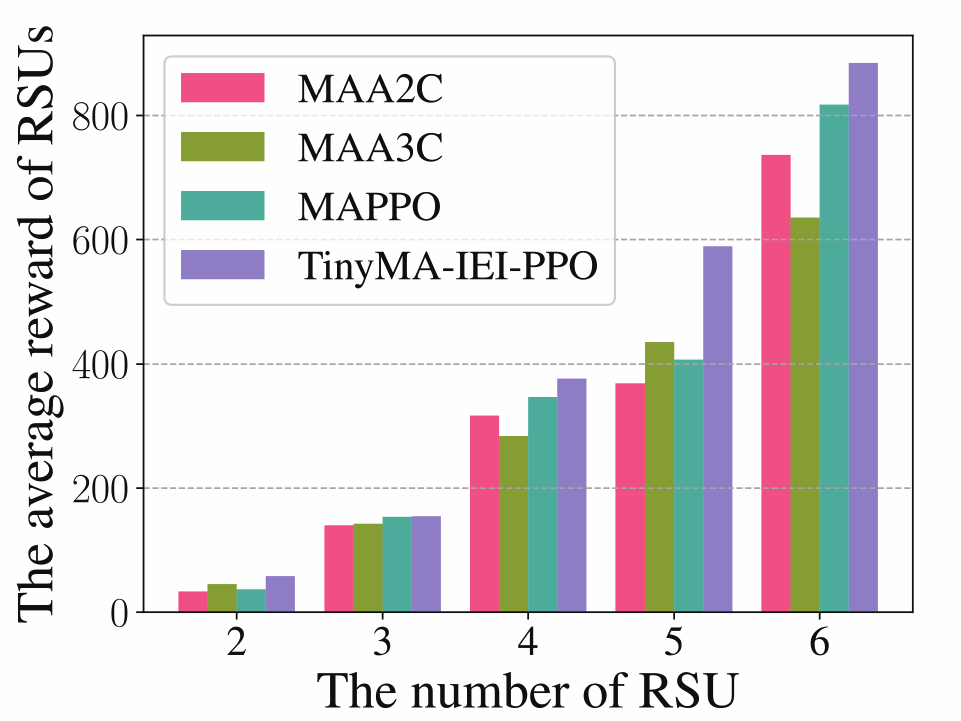}
        \caption{The average reward of RSUs for different numbers of RSUs under different algorithms.}
        \label{RSU_reward_RSU2}
    \end{minipage}
    \label{Reward-Num-change}
\end{figure*}

In this section, we provide numerical results to demonstrate the effectiveness of the proposed approach empirically. We consider 3 AVs and 2 RSUs in the system, the key parameters of the experiment are similar to \cite{nie2020multi} \cite{10505943}. To simulate the situation of limited computational resources allocable on AVs, we conducted our experiments on an NVIDIA Jetson Orin Nano Developer Kit embedded platform. The experiments were run within an Ubuntu 22.04 LTS operating environment using the PyTorch 2.3.0 framework.

Firstly, we demonstrate the convergence of the TinyMA-IMI-PPO algorithm and  compare the alteration in the episode reward of the systerm with five baseline algorithms in Fig.\ref{TinyMA_IEI_PPO_baselines}. The shaded region represents the standard deviation of the average evaluation over 5 runs.The TinyMA-IEI-PPO algorithm converges significantly faster than the other algorithms, though it initially shows large fluctuations in episode rewards due to our exploration mechanism that incentivizes agents to explore behaviors with significant global impacts. Moreover, it most closely approximates the SE value and maintains this approximation.
In contrast, while MAPPO also converges very fast and stably, due to the Gaussian exploration mechanism of the original PPO, it fails to effectively escape from local optimal solutions, resulting in performance inferior to that of TinyMA-IEI-PPO. Baseline algorithms such as MAA2C, MAA3C, Greedy, and Random converge more slowly, with greater fluctuations in episode rewards and larger deviations from the SE.

Fig.\ref{rew_plot} and Fig.\ref{act_plot} respectively illustrate the utilities of AVs and RSUs, as well as the pricing strategies of RSUs and the bandwidth demands of AVs converge after approximately 1000 iterations. This convergence pattern demonstrates the effectiveness of the proposed iterative optimization algorithm in achieving stable resource allocation and pricing decisions in  Vehicular Embodied AI Networks.

From Fig.\ref{AV_reward_AV2} to Fig.\ref{RSU_reward_RSU2}, it can be seen that under different numbers of AVs and RSUs, the performance of the TinyMA-IEI-PPO algorithm proposed in this paper is closest to the theoretical value of SE in terms of the average rewards of RSUs and AVs, indicating its superior effectiveness.

Fig.\ref{AV_reward_AV2} illustrates that with limited RSUs and resources, AVs' average reward drops as their number rises. But social network effects and service complementarity keep the total reward from decreasing sharply.
Fig.\ref{AV_reward_RSU2} illustrates that the average reward of AVs initially exhibits rapid growth but gradually tapers off as the number of RSUs increases. This trend arises because the expanded RSU deployment enriches bandwidth availability in the market, intensifying competition among RSUs and driving down pricing strategies. Consequently, AVs procure ample bandwidth resources to optimize service quality. However, the law of diminishing marginal returns dictates that beyond a certain bandwidth threshold, further allocations yield diminishing improvements in AV rewards. 
Fig.\ref{RSU_reward_AV2}  illustrates that the average reward of RSUs grows very sharply. This is because when resources are relatively limited, RSUs become resource monopolists and set extremely high prices. AVs are in a competitive relationship with each other and each must complete the EAAT migration task, so they are willing to pay high bandwidth fees.
Fig.\ref{RSU_reward_RSU2} reflects that when the bandwidth demand is certain, the more the number of RSUs, the more intense the competition among them. This leads to lower pricing and a decrease in the average reward.

\begin{figure} [ht]
    \centering{
    \includegraphics[width=1\linewidth]{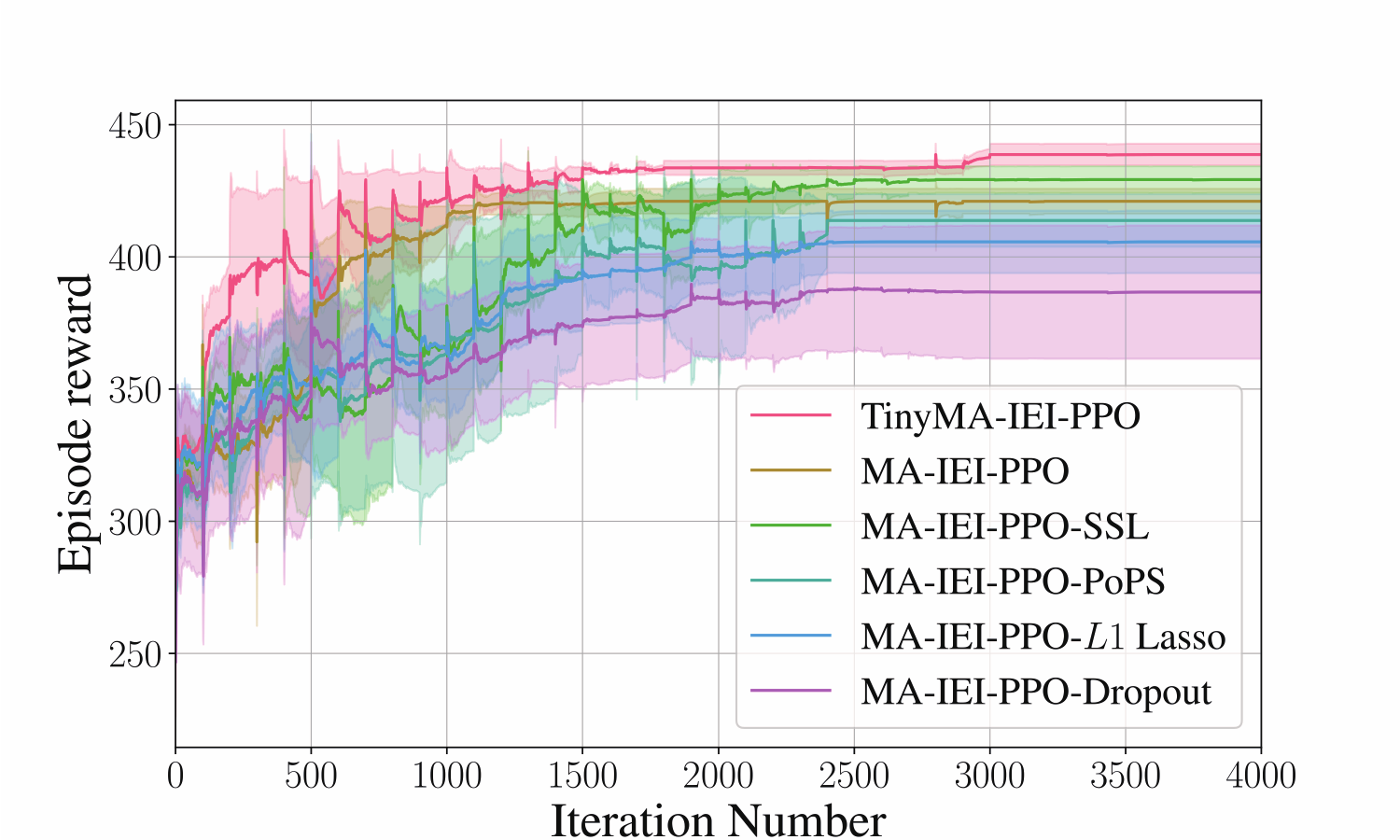}
    \caption{The Episode reward performance of each method with the 85\text{\%} pruning rate}
    \label{Prune_baselines}}
\end{figure}
As shown in Fig.\ref{Prune_baselines}, the self-adaptive dynamic structure pruning method, TinyMA-IEI-PPO demonstrates competitive performance compared to the baseline algorithms. When compared with MA-IEI-PPO, the original algorithm without pruning, TinyMA-IEI-PPO shows that using an appropriate approach to prune redundant neurons can accelerate convergence and enhance the algorithm's performance.MA-IEI-PPO-SSL has a slower convergence rate but improves steadily. Eventually, under extremely sparse conditions (i.e., an 85\text{\&} pruning rate), it can converge stably to a result better than that of the unpruned baseline at around 2500 iterations. MA-IEI-PPO-$L_1$Lasso and MA-IEI-PPO-PoPS have comparable performance. However, under extremely sparse conditions, due to the inherent simplicity of their algorithms, their performance is inferior to that of the unpruned baseline. For MA-IEI-PPO-Dropout, randomly discarding neurons and weights causes significant fluctuations in the algorithm, resulting in poor performance.

\section{Conclusion and Future Work}\label{Conclusion}
In this paper, we introduce EAI-empowered AVs that integrate DTs to generate VEATs and VEAATs, aiming to provide services for AV users. We focused on the scenario where AVs offer seamless in-vehicle services by transferring their VEAATs among RSUs. To achieve efficient migration of VEAATs in VEANETs, we propose a MLMF Stackelberg game-theoretic incentive mechanism. This mechanism incorporates AVs' social influence, service complementarity and substitutability, as well as a virtual immersion index. Additionally, we propose TinyMA-IEI-PPO, a self-adaptive dynamic structured pruning algorithm, to optimize VEAAT migration decisions. Numerical results show that our approach achieves convergence comparable to baseline models and closely approximates the Stackelberg equilibrium. Notably, the TinyMA-IEI-PPO algorithm effectively removes redundant neurons under extremely sparse conditions while maintaining performance, significantly reducing computational overhead. 

For future research, we plan to continue exploring personalized and adaptive pruning algorithms. These algorithms could further optimize computational resources in different scenarios of VEANETs. Additionally, we will investigate alternative modeling approaches beyond the Stackelberg game-theoretic model, such as auction models and the Prospect (PT) theory. We aim to explore whether these models can offer better perspectives for addressing the VEAAT migration problem. Moreover, we will keep integrating the most cutting-edge and advanced technologies with deep reinforcement learning (DRL) to enhance the performance and scalability of our proposed framework in VEANETs.
\bibliographystyle{IEEEtran}

\bibliography{mutil_agent_prune}

\begin{thebibliography}{10}
\providecommand{\url}[1]{#1}
\csname url@samestyle\endcsname
\providecommand{\newblock}{\relax}
\providecommand{\bibinfo}[2]{#2}
\providecommand{\BIBentrySTDinterwordspacing}{\spaceskip=0pt\relax}
\providecommand{\BIBentryALTinterwordstretchfactor}{4}
\providecommand{\BIBentryALTinterwordspacing}{\spaceskip=\fontdimen2\font plus
\BIBentryALTinterwordstretchfactor\fontdimen3\font minus \fontdimen4\font\relax}
\providecommand{\BIBforeignlanguage}[2]{{%
\expandafter\ifx\csname l@#1\endcsname\relax
\typeout{** WARNING: IEEEtran.bst: No hyphenation pattern has been}%
\typeout{** loaded for the language `#1'. Using the pattern for}%
\typeout{** the default language instead.}%
\else
\language=\csname l@#1\endcsname
\fi
#2}}
\providecommand{\BIBdecl}{\relax}
\BIBdecl

\bibitem{savva2019habitat}
M.~Savva, A.~Kadian, O.~Maksymets, Y.~Zhao, E.~Wijmans, B.~Jain, J.~Straub, J.~Liu, V.~Koltun, J.~Malik \emph{et~al.}, ``Habitat: A platform for embodied ai research,'' in \emph{Proceedings of the IEEE/CVF international conference on computer vision}, 2019, pp. 9339--9347.

\bibitem{paolo2024call}
G.~Paolo, J.~Gonzalez-Billandon, and B.~K{\'e}gl, ``A call for embodied ai,'' \emph{arXiv preprint arXiv:2402.03824}, 2024.

\bibitem{sharma2024artificial}
P.~Sharma and C.~Rana, ``Artificial intelligence based object detection and traffic prediction by autonomous vehicles--a review,'' \emph{Expert Systems with Applications}, p. 124664, 2024.

\bibitem{liu2024aligning}
Y.~Liu, W.~Chen, Y.~Bai, X.~Liang, G.~Li, W.~Gao, and L.~Lin, ``Aligning cyber space with physical world: A comprehensive survey on embodied ai,'' \emph{arXiv preprint arXiv:2407.06886}, 2024.

\bibitem{zhong2025generative}
Y.~Zhong, J.~Kang, J.~Wen, D.~Ye, J.~Nie, D.~Niyato, X.~Gao, and S.~Xie, ``Generative diffusion-based contract design for efficient ai twin migration in vehicular embodied ai networks,'' \emph{IEEE Transactions on Mobile Computing}, 2025.

\bibitem{zhang2025embodied}
R.~Zhang, C.~Zhao, H.~Du, D.~Niyato, J.~Wang, S.~Sawadsitang, X.~Shen, and D.~I. Kim, ``Embodied ai-enhanced vehicular networks: An integrated large language models and reinforcement learning method,'' \emph{arXiv preprint arXiv:2501.01141}, 2025.

\bibitem{xiang2023language}
J.~Xiang, T.~Tao, Y.~Gu, T.~Shu, Z.~Wang, Z.~Yang, and Z.~Hu, ``Language models meet world models: Embodied experiences enhance language models,'' \emph{Advances in neural information processing systems}, vol.~36, pp. 75\,392--75\,412, 2023.

\bibitem{li2023intelligent}
W.~Li, D.~Cao, R.~Tan, T.~Shi, Z.~Gao, J.~Ma, G.~Guo, H.~Hu, J.~Feng, and L.~Wang, ``Intelligent cockpit for intelligent connected vehicles: Definition, taxonomy, technology and evaluation,'' \emph{IEEE Transactions on Intelligent Vehicles}, vol.~9, no.~2, pp. 3140--3153, 2023.

\bibitem{chen2024scenario}
H.~Chen, R.~Gao, L.~Fan, E.~Liu, W.~Li, R.~Tan, Y.~Li, L.~He, and D.~Cao, ``Scenario-function system for automotive intelligent cockpits: Framework, research progress and perspectives,'' \emph{IEEE Transactions on Intelligent Vehicles}, 2024.

\bibitem{zhang2023learning}
J.~Zhang, J.~Nie, J.~Wen, J.~Kang, M.~Xu, X.~Luo, and D.~Niyato, ``Learning-based incentive mechanism for task freshness-aware vehicular twin migration,'' in \emph{2023 IEEE 43rd International Conference on Distributed Computing Systems Workshops (ICDCSW)}.\hskip 1em plus 0.5em minus 0.4em\relax IEEE, 2023, pp. 103--108.

\bibitem{chen2023multiagent}
J.~Chen, J.~Kang, M.~Xu, Z.~Xiong, D.~Niyato, C.~Chen, A.~Jamalipour, and S.~Xie, ``Multiagent deep reinforcement learning for dynamic avatar migration in aiot-enabled vehicular metaverses with trajectory prediction,'' \emph{IEEE Internet of Things Journal}, vol.~11, no.~1, pp. 70--83, 2023.

\bibitem{nie2020multi}
J.~Nie, J.~Luo, Z.~Xiong, D.~Niyato, P.~Wang, and H.~V. Poor, ``A multi-leader multi-follower game-based analysis for incentive mechanisms in socially-aware mobile crowdsensing,'' \emph{IEEE Transactions on Wireless Communications}, vol.~20, no.~3, pp. 1457--1471, 2020.

\bibitem{li2019stackelberg}
F.~Li, H.~Yao, J.~Du, C.~Jiang, and Y.~Qian, ``Stackelberg game-based computation offloading in social and cognitive industrial internet of things,'' \emph{IEEE Transactions on Industrial Informatics}, vol.~16, no.~8, pp. 5444--5455, 2019.

\bibitem{Xu_Peng_Gupta_Kang_Xiong_Li_El-Latif_2022}
\BIBentryALTinterwordspacing
M.~Xu, J.~Peng, B.~B. Gupta, J.~Kang, Z.~Xiong, Z.~Li, and A.~A.~A. El-Latif, ``\BIBforeignlanguage{en-US}{Multiagent federated reinforcement learning for secure incentive mechanism in intelligent cyber–physical systems},'' \emph{\BIBforeignlanguage{en-US}{IEEE Internet of Things Journal}}, p. 22095–22108, Nov 2022. [Online]. Available: \url{http://dx.doi.org/10.1109/jiot.2021.3081626}
\BIBentrySTDinterwordspacing

\bibitem{disabato2022tiny}
S.~Disabato and M.~Roveri, ``Tiny machine learning for concept drift,'' \emph{IEEE Transactions on Neural Networks and Learning Systems}, 2022.

\bibitem{song2023comprehensive}
Y.~Song, T.~Wang, P.~Cai, S.~K. Mondal, and J.~P. Sahoo, ``A comprehensive survey of few-shot learning: Evolution, applications, challenges, and opportunities,'' \emph{ACM Computing Surveys}, vol.~55, no. 13s, pp. 1--40, 2023.

\bibitem{loftin2021strategically}
R.~Loftin, A.~Saha, S.~Devlin, and K.~Hofmann, ``Strategically efficient exploration in competitive multi-agent reinforcement learning,'' in \emph{Uncertainty in Artificial Intelligence}.\hskip 1em plus 0.5em minus 0.4em\relax PMLR, 2021, pp. 1587--1596.

\bibitem{yang2024embodied}
Y.~Yang, Y.~Chen, J.~Wang, G.~Sun, and D.~Niyato, ``Embodied ai-empowered low altitude economy: Integrated sensing, communications, computation, and control (isc3),'' \emph{arXiv preprint arXiv:2412.19996}, 2024.

\bibitem{zhou2024embodied}
Y.~Zhou, L.~Huang, Q.~Bu, J.~Zeng, T.~Li, H.~Qiu, H.~Zhu, M.~Guo, Y.~Qiao, and H.~Li, ``Embodied understanding of driving scenarios,'' \emph{arXiv preprint arXiv:2403.04593}, 2024.

\bibitem{10608164}
Y.~Tong, J.~Chen, M.~Xu, J.~Kang, Z.~Xiong, D.~Niyato, C.~Yuen, and Z.~Han, ``Multi-attribute auction-based resource allocation for twins migration in vehicular metaverses: A gpt-based drl approach,'' \emph{IEEE Transactions on Cognitive Communications and Networking}, vol.~11, no.~1, pp. 638--654, 2025.

\bibitem{10302973}
J.~Zhang, J.~Nie, J.~Wen, J.~Kang, M.~Xu, X.~Luo, and D.~Niyato, ``Learning-based incentive mechanism for task freshness-aware vehicular twin migration,'' in \emph{2023 IEEE 43rd International Conference on Distributed Computing Systems Workshops (ICDCSW)}, 2023, pp. 103--108.

\bibitem{kang2024tiny}
J.~Kang, Y.~Zhong, M.~Xu, J.~Nie, J.~Wen, H.~Du, D.~Ye, X.~Huang, D.~Niyato, and S.~Xie, ``Tiny multiagent drl for twins migration in uav metaverses: A multileader multifollower stackelberg game approach,'' \emph{IEEE Internet of Things Journal}, vol.~11, no.~12, pp. 21\,021--21\,036, 2024.

\bibitem{10505943}
J.~Kang, J.~Zhang, H.~Yang, D.~Ye, and M.~S. Hossain, ``When metaverses meet vehicle road cooperation: Multiagent drl-based stackelberg game for vehicular twins migration,'' \emph{IEEE Internet of Things Journal}, vol.~11, no.~22, pp. 35\,928--35\,941, 2024.

\bibitem{10734312}
R.~Wang, Y.~Jing, C.~Gu, S.~He, and J.~Chen, ``End-to-end multitarget flexible job shop scheduling with deep reinforcement learning,'' \emph{IEEE Internet of Things Journal}, vol.~12, no.~4, pp. 4420--4434, 2025.

\bibitem{kim2024strangeness}
J.-B. Kim, H.-B. Choi, and Y.-H. Han, ``Strangeness-driven exploration in multi-agent reinforcement learning,'' \emph{Neural Networks}, vol. 172, p. 106149, 2024.

\bibitem{li2024individual}
X.~Li, Z.~Liu, S.~Chen, and J.~Zhang, ``Individual contributions as intrinsic exploration scaffolds for multi-agent reinforcement learning,'' \emph{arXiv preprint arXiv:2405.18110}, 2024.

\bibitem{zheng2021episodic}
L.~Zheng, J.~Chen, J.~Wang, J.~He, Y.~Hu, Y.~Chen, C.~Fan, Y.~Gao, and C.~Zhang, ``Episodic multi-agent reinforcement learning with curiosity-driven exploration,'' \emph{Advances in Neural Information Processing Systems}, vol.~34, pp. 3757--3769, 2021.

\bibitem{zhang2023self}
S.~Zhang, J.~Cao, L.~Yuan, Y.~Yu, and D.-C. Zhan, ``Self-motivated multi-agent exploration,'' \emph{arXiv preprint arXiv:2301.02083}, 2023.

\bibitem{zhou2023online}
F.~Zhou, X.~Qiu, Z.~Cai, W.~Chen, H.~Zhao, and Z.~Li, ``Online-s2t: A lightweight distributed online reinforcement learning training framework for resource-constrained devices,'' in \emph{2023 Asia Conference on Advanced Robotics, Automation, and Control Engineering (ARACE)}.\hskip 1em plus 0.5em minus 0.4em\relax IEEE, 2023, pp. 93--101.

\bibitem{livne2020pops}
D.~Livne and K.~Cohen, ``Pops: Policy pruning and shrinking for deep reinforcement learning,'' \emph{IEEE Journal of Selected Topics in Signal Processing}, vol.~14, no.~4, pp. 789--801, 2020.

\bibitem{su2024compressing}
W.~Su, Z.~Li, M.~Xu, J.~Kang, D.~Niyato, and S.~Xie, ``Compressing deep reinforcement learning networks with a dynamic structured pruning method for autonomous driving,'' \emph{IEEE Transactions on Vehicular Technology}, 2024.

\bibitem{CHRISLEY2003131}
\BIBentryALTinterwordspacing
R.~Chrisley, ``Embodied artificial intelligence,'' \emph{Artificial Intelligence}, vol. 149, no.~1, pp. 131--150, 2003. [Online]. Available: \url{https://www.sciencedirect.com/science/article/pii/S0004370203000559}
\BIBentrySTDinterwordspacing

\bibitem{10352433}
G.~C. Alexandropoulos, N.~Shlezinger, I.~Alamzadeh, M.~F. Imani, H.~Zhang, and Y.~C. Eldar, ``Hybrid reconfigurable intelligent metasurfaces: Enabling simultaneous tunable reflections and sensing for 6g wireless communications,'' \emph{IEEE Vehicular Technology Magazine}, vol.~19, no.~1, pp. 75--84, 2024.

\bibitem{wang2025pre}
Y.~Wang, Y.~Mei, Z.~Gao, Z.~Wan, B.~Ning, D.~Mi, and S.~Muhaidat, ``Pre-equalization aided grant-free massive access in massive mimo system,'' \emph{arXiv preprint arXiv:2502.06239}, 2025.

\bibitem{Huang_Zhong_Nie_Hu_Xiong_Kang_Quek_2022}
X.~Huang, W.~Zhong, J.~Nie, Q.~Hu, Z.~Xiong, J.~Kang, and T.~Quek, ``\BIBforeignlanguage{en-US}{Joint user association and resource pricing for metaverse: Distributed and centralized approaches},'' Aug 2022.

\bibitem{ding2021locally}
K.~Ding, Y.~Liu, X.~Zou, S.~Wang, and K.~Ma, ``Locally adaptive structure and texture similarity for image quality assessment,'' in \emph{Proceedings of the 29th ACM International Conference on multimedia}, 2021, pp. 2483--2491.

\bibitem{yu2024attention}
J.~Yu, A.~Alhilal, T.~Zhou, P.~Hui, and D.~H. Tsang, ``Attention-based qoe-aware digital twin empowered edge computing for immersive virtual reality,'' \emph{IEEE Transactions on Wireless Communications}, 2024.

\bibitem{Du_Liu_Niyato_Kang_Xiong_Zhang_Kim_2022}
H.~Du, J.~Liu, D.~Niyato, J.~Kang, Z.~Xiong, J.~Zhang, and D.~Kim, ``\BIBforeignlanguage{en-US}{Attention-aware resource allocation and qoe analysis for metaverse xurllc services},'' Aug 2022.

\bibitem{415dc80a-289c-39e1-8e6b-601fc5ef267e}
\BIBentryALTinterwordspacing
R.~B. MYERSON, \emph{Game Theory: Analysis of Conflict}.\hskip 1em plus 0.5em minus 0.4em\relax Harvard University Press, 1991. [Online]. Available: \url{http://www.jstor.org/stable/j.ctvjsf522}
\BIBentrySTDinterwordspacing

\bibitem{Xu_Qiu_Zhang_Liu_Liu_Chen_2021}
\BIBentryALTinterwordspacing
H.~Xu, X.~Qiu, W.~Zhang, K.~Liu, S.~Liu, and W.~Chen, ``\BIBforeignlanguage{en-US}{Privacy-preserving incentive mechanism for multi-leader multi-follower iot-edge computing market: A reinforcement learning approach},'' \emph{\BIBforeignlanguage{en-US}{Journal of Systems Architecture}}, p. 101932, Mar 2021. [Online]. Available: \url{http://dx.doi.org/10.1016/j.sysarc.2020.101932}
\BIBentrySTDinterwordspacing

\end{thebibliography}

\end{document}